\documentclass[12pt]{article}
\usepackage{epsfig}
\usepackage{latexsym}

\newcommand{\figwidth}{12.5cm}

\newcommand{\gapprox}{{\raisebox{-0.5ex}{${\scriptstyle>}$} \atop \raisebox{0.5ex}{${\scriptstyle\sim}$}}}
\newcommand{\lapprox}{{\raisebox{-0.5ex}{${\scriptstyle<}$} \atop \raisebox{0.5ex}{${\scriptstyle\sim}$}}}
\newcommand{\boxed}[1]{\setlength{\fboxsep}{1pt}\fbox{#1}}

\title{Quark phases in neutron stars and a \mbox{``third family''} of compact stars \\
as a signature for phase transitions$^{}$\footnote{Supported by DFG and GSI Darmstadt.}}
\author{K.~Schertler$^a$, 
C.~Greiner$^a$, J. Schaffner-Bielich$^b$, and M.H.~Thoma$^{c,}$\footnote{Heisenberg Fellow.}}
\date{}
\begin{document}
\maketitle
\begin{center}
{\small \it $^a$Institut f\"ur Theoretische Physik, Universit\"at Giessen \\
35392 Giessen, Germany }

\smallskip
{\small \it $^b$RIKEN BNL Research Center, Brookhaven National Laboratory\\
Upton, New York 11973-5000, USA}

\smallskip
{\small \it $^c$Theory Division CERN, CH-1211 Geneva 23, Switzerland}

\end{center}

\begin{figure}[ht]
\centerline{\epsfig{file=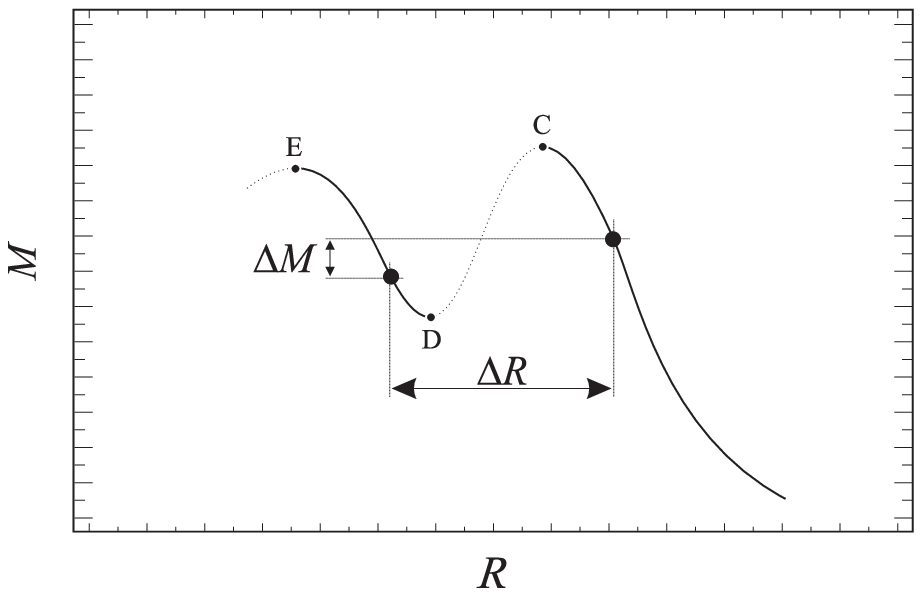,height=4.5cm}}
\end{figure}

\begin{abstract}
The appearance of quark phases in the dense interior of neutron stars provides one possibility 
to soften the equation of state (EOS) of neutron star matter at high densities. 
This softening leads to more compact equilibrium configurations of neutron stars 
compared to pure hadronic stars of the same mass. We investigate the question to which amount
the compactness of a neutron star can be attributed to the presence of a quark phase.
For this purpose we employ several hadronic EOS in the framework of the relativistic mean-field 
(RMF) model and an extended MIT bag model to describe the quark phase. 
We find that - almost independent of the model parameters - the radius of a pure hadronic neutron star 
gets typically reduced by $20-30\%$ if a pure quark phase in the center of the star does exist.
For some EOS we furthermore find the possibility of a {\em third family} of compact stars 
which may exist besides the two known families of white dwarfs and neutron stars. 
We show how an experimental proof of
the existence of a third family by mass and radius measurements may provide a unique signature
for a phase transition inside neutron stars.


\end{abstract}

\section{Introduction}

The properties of nuclear matter under extreme conditions are subject to intense experimental and 
theoretical studies in heavy-ion physics as well as in nuclear astrophysics.
Especially neutron stars provide a unique astrophysical environment to study the properties of matter
in the region of low temperatures and high baryonic densities.
The densities deep inside neutron stars might be sufficient to populate numerous new forms of matter
which may compete with each other. 
It is expected that at about two times normal nuclear matter density hyperons appear 
in neutron star matter \cite{GlenBook,WeberBook,SchaMish96,Balb98}.
Also the formation of Bose-Einstein condensates \cite{Glen85,GlenScha98,Kaon} 
or the phase transition to deconfined quark matter have been considered 
\cite{GlenBook,WeberBook,BaymChin76,Quarks,MH,Sche98,Sche99}. 
For a recent review on possible phase transitions
in neutron stars see \cite{HeisHjor99}.
However, our understanding of the properties of neutron star matter is still restricted due to 
theoretical limits and uncertainties in modeling the equation of state (EOS).
Measurements of neutron star masses and radii may put important 
constraints on the EOS at large density.
The masses of about twenty neutron stars (mainly pulsars in binary systems) have been
measured and found to be quite near the ``canonical'' mass of $M=1.4\,M_\odot$ 
($M_\odot$ is the solar mass) \cite{Thor}. 
Unfortunately, the currently known masses do not constrain the EOS unambiguously.
While some soft EOS -- which can only account for smaller masses -- can be ruled out, 
it is not possible to differentiate between the class of stiffer EOS which give 
higher neutron star masses. However, 
these EOS can be classified by their different predictions for neutron star radii. 
Precise measurements of neutron star radii can therefore give important information on
the property of matter and possible exotic phases at large density.
Some indirect radius estimates from observations of neutron stars 
seem to indicate surprisingly small radii. 
Recent radius estimates for Geminga (PSR 0630+17) by Golden and Shearer \cite{GoldShea99} 
suggest an upper bound 
of the apparent radius of $R_\infty \le 9.5^{+3.5}_{-2.0}\,$km for an assumed blackbody source
and $R_\infty \le 10.0^{+3.8}_{-2.1}\,$km with the presence of a magnetized $H$ atmosphere. 
Walter and Matthews obtained $R_\infty\lapprox 14\,$km for the
recently discovered isolated neutron star RXJ1856.5-3745 \cite{Walt9697}.
Generally the apparent radius $R_\infty$ at a distance $d\rightarrow\infty$ is related to the 
local stellar radius $R$ by
\begin{equation}\label{Rinf}
R_\infty = R/\sqrt{1-2 G M/R c^2}.
\end{equation} 
If we assume a mass of $M=1.4\,M_\odot$ an apparent radius $R_\infty$ of about $11-13\,$km
would restrict the radius to the quite low values of $R \approx 7\!-\!10\,$km. (Due to (\ref{Rinf}), radii
below $R_\infty \approx 11\,$km would imply masses below $1.4\,M_\odot$.)
A radius of $R \lapprox 10\,$km for a typical neutron star mass 
is furthermore consistent with the findings by Haberl and Titarchuk \cite{HabeTita95} for 
the X-ray burster 4U\,1820-30, with the findings by Pavlov and Zavlin \cite{PavlZavl97} for 
the millisecond pulsar J0437-4715 and by Li et al. \cite{Li95} 
for Her X-1. (In the context of Ref.~\cite{Li95} see also Ref. \cite{WassShap83}
and Refs. \cite{Reyn97,Mads97}.)
If radius estimates of $R \lapprox 10\,$km indeed should prove to be reliable in connection with
typical neutron star masses, the careful interpretation of these results
in terms of neutron star matter properties would be an intriguing challenge for nuclear astrophysics.
Since the appearance of a deconfined quark matter phase inside neutron stars is one possible
mechanism to soften the EOS (and therefore to reduce the radius of the star), we want 
systematically address the question to
which amount the compactness of a neutron star can be attributed to the presence of a quark phase.
Such investigations are necessary also in view of more ``exotic'' 
interpretations of small radii e.g.~in terms of ``strange stars'' \cite{Li95,Bomb97,DeyBomb98}. 
It is important to study whether such interpretations are a  
consequence of ruling out other mechanisms suitable to soften the EOS at high densities
(e.g.~by hyperons, kaon condensates or quark phases) or provide only one 
possible explanation of the observed neutron star properties. 
For this purpose we will apply various hadronic EOS in the relativistic mean-field model 
(including hyperons) and a wide range of model parameters of the quark phase 
which we will describe in an extended MIT bag model. 

The EOS of the hadronic and the quark phase will be discussed in Sec.~\ref{EOS}. 
In Sec.~\ref{Phasetransition} we discuss the construction of a first-order 
phase transition from hadronic to quark matter and present the results for the EOS. 
In Sec.~\ref{StarStructure} we apply the EOS to the calculation of the neutron
star structure and discuss the influence of quark phases. In Sec.~\ref{Twins} we 
discuss the properties of a possible ``third family'' of compact stars.
Such hypothetical family of compact stars might exist besides the two 
known families of white dwarfs and neutron
stars \cite{Gerl68,GerlachPhD,GlenKett98}. 
In this work we show how future measurements of masses and radii 
of only two compact stars may prove the existence of a third family. 
Moreover, we argue that this would provide a novel signature for a 
phase transition inside neutron stars, 
revealing a particular behavior of the EOS in the almost unknown regime of supernuclear densities.
Finally, we summarize our findings in Sec.~\ref{Summary}.

\section{Equations of state}\label{EOS}
We want to start with a discussion of the different models used to describe the equation
of state of ``neutron star matter'' i.e.~cold, charge neutral matter in beta equilibrium. Neutron
star matter covers a wide range of densities. Starting from the density of iron 
($\epsilon \approx8 \,$g/cm$^3$) which builds the surface of a neutron star, densities 
up to few times normal nuclear matter density \mbox{($\epsilon_0 = 140\,$MeV/fm$^3$}
\mbox{$\approx 2.5 \times 10^{14}\,$g/cm$^3$)} can be achieved
in the center of the star. 
Unfortunately there is yet no single theory capable to meet the demands of the 
various degrees of freedom opened up in neutron star matter in different densities regimes.
We are therefore forced to use different models at different density ranges.

\subsection{Subnuclear densities\label{SectBPS}}
For subnuclear densities we use the Baym-Pethick-Sutherland 
EOS \cite{BPS}, which describes the crust of the neutron star. Up to densities of 
$\epsilon \,\lapprox 10^7\,$g/cm$^3$ matter is composed of a Coulomb lattice 
of $^{56}_{28}$Fe nuclei. The pressure is dominated by degenerate electrons.
At higher densities nuclei become more and more neutron rich and at 
$\epsilon_{drip}\approx$ \mbox{$4\times 10^{11}\,$g/cm$^3$} the most weakly bound neutrons 
start to drip out of the nuclei.
For a detailed discussion of the Baym-Pethick-Sutherland EOS see \cite{ShapiroBook}.
A review of the properties of neutron rich matter at subnuclear densities can be found in 
\cite{PethRave95}.

\subsection{Nuclear densities\label{SectRMF}}
The Baym-Pethick-Sutherland EOS is matched at $\epsilon \approx 10^{14}\,$g/cm$^3\approx \epsilon_0$ 
to a hadronic EOS calculated in the framework of the relativistic mean-field (RMF) model. 
At these densities nuclei begin to dissolve and nucleons become the 
relevant degrees of freedom in this phase.
The RMF model is widely used for the description of dense nuclear matter 
\cite{RMF,Glen8287,SchaMish96}, especially in neutron 
stars. For an introduction into the RMF model see e.g.\,\cite{GlenBook}.
We use three EOS calculated by Schaffner and Mishustin in the extended RMF model 
\cite{SchaMish96} (i.e.~TM1, TM2, GL85) and one by Ghosh, Phatak and Sahu 
\cite{Ghosh95}. For the latter one we use GPS as an abbreviation. These models include 
hyperonic degrees of freedom which typically appear at $\epsilon \approx 2\!-\! 3 \epsilon_0$. 
Table \ref{NuclProps} shows the
nuclear matter properties and the particle composition of the four EOS. 
%
\begin{table}[ht]
\begin{center}
\begin{tabular}{ccccc}
\hline
Hadronic EOS &  TM1 & TM2 & GL85 & GPS \\
\hline
reference & \cite{SchaMish96} & \cite{SchaMish96} & \cite{SchaMish96} & \cite{Ghosh95} \\  
$\rho_0\,$[fm$^{-3}$] & $0.145$ & $0.132$ & $0.145$ & $0.150$ \\ 
$B/A\,$[MeV] & $-16.3$ & $-16.2$ & $-15.95$ & $-16.0$ \\
$K\,$[MeV] & $281$ & $344$ & $285$ & $300$ \\
$m_N^*/m_n$ & $0.634$ & $0.571$ & $0.770$ & $0.830$ \\
$a_{sym}\,$[MeV] & $36.9$ & $35.8$ & $36.8$ & $32.5$ \\
composition & a & a & a & b \\
\hline
\multicolumn{5}{l}{a) n, p, e$^-$, $\mu^-$, $\Lambda$, $\Sigma^-$, $\Sigma^0$, $\Sigma^+$,
$\Xi^-$, $\Xi^0$}\\
\multicolumn{5}{l}{b) n, p, e$^-$, $\mu^-$, $\Lambda$, $\Sigma^-$}\\
\hline
\end{tabular}
\end{center}
\caption{Nuclear matter properties of the hadronic EOS. The saturation density and
the binding energy is denoted by $\rho_0$ and $B/A$, the incompressibility by $K$, the
effective mass by $m_N^*/m_n$ and the symmetry energy by $a_{sym}$. The particle compositions
are shown at the bottom of the table.}
\label{NuclProps}
\end{table}
%
Even if the relevant degrees of freedom are specified
(in our case basically nucleons and hyperons) the high density range of the EOS is still not well 
understood. The use of different hadronic models should reflect this uncertainty to some degree.
In the following we will denote the phase described by the Baym-Pethick-Sutherland EOS 
and by the RMF model as the {\em hadronic phase} (HP) of the neutron star.

\subsection{Quark matter\label{SectQP}}
At densities above $\epsilon_0$ we allow the HP to undergo a phase transition to a deconfined
{\em quark matter phase} (QP). This section describes how we model the QP in weak
equilibrium. The QP consists of $u$, $d$, $s$ quarks and electrons in weak
equilibrium i.e.~the weak reactions
\begin{eqnarray}
d & \longrightarrow & u+e^- +\bar{\nu}_{e^-}, \\
s & \longrightarrow & u+e^- +\bar{\nu}_{e^-}, \nonumber \\
s + u & \longleftrightarrow & d + u, \nonumber
\end{eqnarray}
imply relations between the four chemical potentials $\mu_u$, $\mu_d$, $\mu_s$, $\mu_e$ which read
\begin{equation}
\mu_s  = \mu_d =  \mu_u+\mu_e.
\end{equation}
Since the neutrinos can diffuse out of the star their chemical potentials are taken to be zero.
The number of chemical potentials necessary for the description of the QP in weak
equilibrium (the number of components)
is therefore reduced to two independent ones.  We choose the pair ($\mu_n$, $\mu_e$)
with the neutron chemical potential
\[\mu_n \equiv \mu_u + 2 \mu_d. \]
In a pure QP (in contrast to a QP in a mixed phase, which we will discuss later) we have to 
require the QP to be charge neutral. This gives us an additional constraint on the chemical potentials
\begin{equation}
\rho_c^{QP}=\frac{2}{3} \rho_u -\frac{1}{3} \rho_d 
-\frac{1}{3} \rho_s -\rho_e=0.
\end{equation}
Here $\rho_c^{QP}$ denotes the charge density of the QP and 
$\rho_f$ ($f\in u,d,s$), $\rho_e$ the particle densities of the quarks and the electrons, respectively.
The EOS can now be parametrized by only one chemical potential, say $\mu_n$. At this point it should
be noted that the arguments given here for the QP also holds for the HP. There one also ends up
with two independent chemical potentials (e.g.~$\mu_n$, $\mu_e$) if one only requires weak equilibrium
between the constituents of the HP, and
with one chemical potential (e.g.~$\mu_n$) if one additionally requires charge neutrality. As we will 
discuss later, the number of independent chemical potentials plays a crucial role in the formulation
of the Gibbs condition for chemical and mechanical equilibrium between the HP and the QP. 

To calculate the EOS of the QP we apply the ``effective mass bag model'' \cite{Sche98,Sche97}.
In this model, medium effects are taken into account in the framework of the MIT bag model 
\cite{MITbag,FarhiJaffe84} by introducing  density-dependent effective quark masses. 
%
\begin{figure}[hb]
\centerline{\epsfig{file=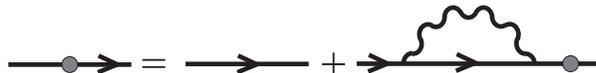,height=1cm}}
\caption{Resummed quark propagator.}
\label{oneloop}
\end{figure}
%
The effective quark masses follow from the poles of the resummed one-loop quark propagator at finite 
chemical potential (see Fig.~\ref{oneloop}).
Details of the calculation of the effective quark masses in the so called hard dense loop approximation
can be found in \cite{Sche97}. 
One finds \cite{Sche97,EffeMass}
\begin{equation} \label{defmeff}
m_f^*(\mu_f)=\frac{m_f}{2}+\sqrt {\frac{m_f^2}{4}+\frac{g^2\mu_f^2}{6\pi ^2}},
\end{equation}
with the current quark mass $m_f$ of the quark flavor $f$ and the strong coupling constant $g$.
These effective quark masses show the behavior of increasing mass with increasing 
density (chemical potential). For the current quark masses we choose $m_u=m_d=0$ and 
$m_s=150\,$MeV.
The effective masses (\ref{defmeff}) can now be used in the Fermi gas expression for the particle density
\begin{equation}
  \rho_f(\mu_f) = \frac{d_f}{6 \pi^2} \left[\mu_f^2-{m_f^*}^2(\mu_f)\right]^{3/2},
\end{equation}
where $d_f$ is the degree of degeneracy \cite{GoreYang95}. 
The pressure $p_f$ and energy density $\epsilon_f$ can be extracted from the thermodynamic relations
\begin{equation}
        \rho_f(\mu_f) = \frac{d p_f(\mu_f)}{d \mu_f} \quad \mbox{and} \quad 
        \epsilon_f(\mu_f)+p_f(\mu_f) = \mu_f \rho_f(\mu_f).
\end{equation}
If $g \rightarrow 0$ the effective quark masses (\ref{defmeff}) approach the current quark mass 
\mbox{$m_f$}. In this limit, the EOS equals to a free MIT bag Fermi gas EOS. 
We therefore can use $g$ to study the deviations from the free Fermi gas owing to
the influence of medium effects. For this purpose we take $g$ as a parameter ranging from
$g=0$ (no medium effects) to $g \approx 2-3$ ($\alpha_s \approx 0.31 -0.72$).
(It was found in \cite{Sche98} that a phase transition 
to a pure QP does not occur inside a typical neutron star for $g \gapprox 2$. 
Larger values are therefore of minor interest for our purpose of studying 
the influence of quark phases on the properties of neutron stars.) 
The overall energy density $\epsilon_{QP}$ and pressure $p_{QP}$ 
of the QP is the sum over all flavors plus the Fermi gas contribution 
$\epsilon_e$, $p_e$ of the uniform background of electrons plus the contribution 
of the phenomenological bag constant $B$.
\begin{eqnarray} \label{QPepsp}
\epsilon_{QP} & = & \epsilon_u+\epsilon_d+\epsilon_s+\epsilon_e +B, \\
p_{QP} & = & p_u +p_d +p_s +p_e -B, \nonumber
\end{eqnarray}
with 
\begin{equation}
\epsilon_e = \frac{\mu_e^4}{4 \pi^2} \quad \mbox{and} \quad p_e = \frac{\mu_e^4}{12 \pi^2}.
\end{equation}
The bag constant $B$ is introduced in the usual way by adding it to the 
energy density and subtracting it from the pressure. 
Due to the phenomenological nature of the bag model, an exact value of $B$ is
not known. Therefore we will take $B$ as a free parameter 
ranging from $B^{1/4} = 165\,$MeV ($B \approx 96\,$MeV/fm$^3$) to $B^{1/4} \approx 200\,$MeV 
($B \approx 208\,$MeV/fm$^3$). The lower bound of the bag constant is reasonable if one
requires that the deconfinement phase transition to the QP should not occur at densities below
$\epsilon \approx \epsilon_0$. 
We do not consider here the possibility of the existence of pure QP stars, so called ``strange stars'' 
or the related case of an absolutely stable QP (stable strange quark matter) with respect to the HP
as it was supposed by Witten \cite{Bodm71Witte84}. These possibilities are in detail discussed in the
literature and we refer to \cite{GlenBook,WeberBook,Mads98,GreiScha} for an overview.
To find the QP to be absolutely stable, their energy per baryon
$E/A$ in equilibrium ($p_{QP}=0$) must be lower than the energy per baryon of  $^{56}$Fe
which is about $930\,$MeV. This requires small bag constants of $B^{1/4} \lapprox 155\,$MeV
(if we neglect medium effects). At our lower bound of the bag constant of $B^{1/4} = 165\,$MeV we
obtain an energy per baryon of $E/A \approx 990\,$MeV which is about $60\,$MeV unbound with
respect to $^{56}$Fe.
As one can imagine from (\ref{QPepsp}), the bag constant plays a crucial role in the question 
whether or not a QP can exist in the interior of neutron stars.

\section{Phase transition}\label{Phasetransition}
Using the four EOS of the HP discussed in section \ref{SectBPS} and \ref{SectRMF} and the 
EOS of the QP discussed in the last section (\ref{SectQP}), we study the
deconfinement phase transition from the HP to the QP and calculate the corresponding total EOS. 
In particular we are interested in the influence
of the model parameters of the QP (i.e.~the bag constant $B$ and coupling constant $g$) on the 
phase transition densities. (The existence of a QP inside the neutron star clearly requires
the transition density to be smaller than the central density of the star.) Before we come to this, 
we have first to discuss how to construct a first-order phase transition from the confined 
HP to the deconfined QP.

\subsection{Construction of the phase transition}

\begin{figure}[ht]
\centerline{\epsfig{file=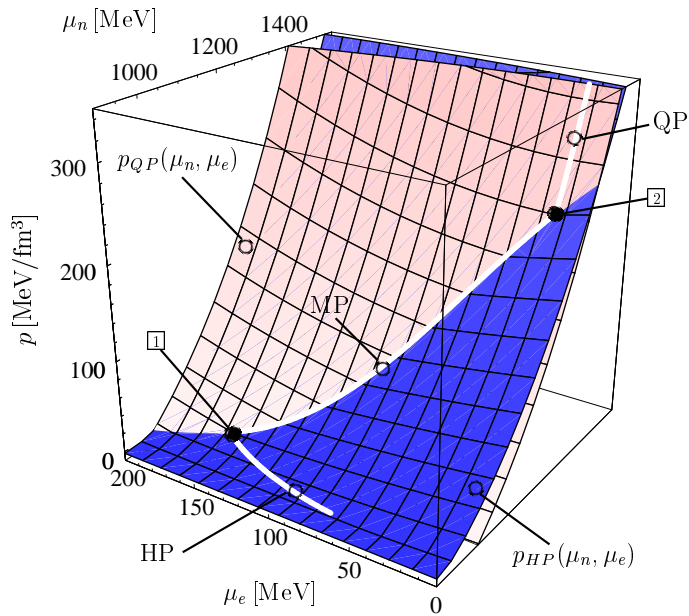,width=9cm}}
\setlength{\fboxsep}{2pt}
\caption{Gibbs phase construction of a two-component system. Plotted is the pressure of the hadronic
phase $p_{HP}$ and the pressure of the quark phase $p_{QP}$ as a function of the two independent
chemical potentials $\mu_n$, $\mu_e$. The white lines HP and QP on the pressure surfaces show the
pressure of the hadronic phase and the quark phase under the condition of charge neutrality. 
At low pressure matter is in its charge neutral HP. 
At point \boxed{1} the HP reaches the intersection curve MP of the mixed phase. This curve is
the solution of the Gibbs condition (\ref{GibbsCondition}). 
At a pressure above point \boxed{2} matter consists of a pure QP.
EOS of the HP is GPS, EOS of the QP uses $B^{1/4}=170\,$MeV and $g=2$.}
\label{p3D}
\end{figure}

\begin{figure}[th]
\centerline{\epsfig{file=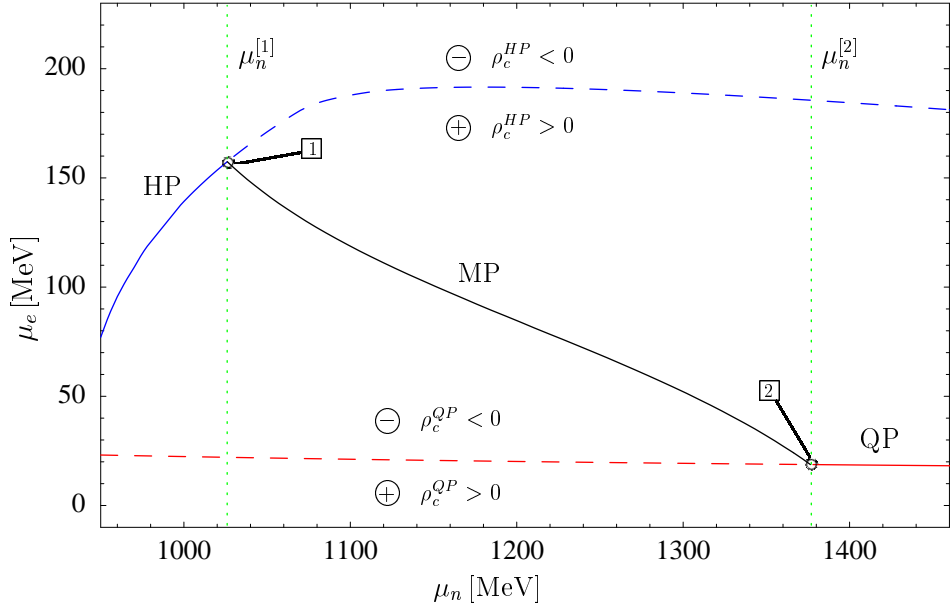,width=\figwidth}}
\setlength{\fboxsep}{2pt}
\caption{Electron chemical potential $\mu_e$ as a function of the neutron chemical potential $\mu_n$.
Up to $\mu_n^{[1]}$ matter is in its charge neutral HP. From $\mu_n^{[1]}$ to $\mu_n^{[2]}$ matter consists
of a mixed phase (MP) of a charge positive HP 
($\rho_c^{HP}>0$, $\rho_c$ denotes the charge density) and a charge negative QP ($\rho_c^{QP}<0$).
The MP curve is calculated from the Gibbs condition (\ref{GibbsCondition}).
Above $\mu_n^{[2]}$ matter consists of a pure charge neutral QP. 
EOS of the HP is GPS, EOS of the QP uses $B^{1/4}=170\,$MeV and $g=2$.}
\label{munmue}
\end{figure}

We calculate the phase transition according to Glendenning who has first
realized the possibility of the occurrence of a 
mixed phase (MP) of hadronic and quark matter in a finite density
range inside neutron stars \cite{GlenBook,Glen92}. For simplicity
we neglect Coulomb and surface effects in the MP as studied in Ref.~\cite{Heise93}.  
The essential point in the calculation of the MP is that total charge neutrality can be achieved
by a positively charged amount of hadronic matter and a negatively charged amount of 
quark matter. 
As already discussed in Sec.~\ref{SectQP} we have to deal with two 
independent chemical potentials ($\mu_n$, $\mu_e$) if we impose the condition of weak equilibrium.
Such a system is called a {\em two-component system}. 
The Gibbs condition for mechanical and chemical equilibrium at zero temperature between
the both phases reads
\begin{equation}\label{GibbsCondition}
p_{HP}(\mu_n, \mu_e) = p_{QP}(\mu_n, \mu_e).
\end{equation}
Using Eq.~(\ref{GibbsCondition}) we can calculate the equilibrium chemical potentials of the MP where 
$p_{HP}=p_{QP} \equiv p_{MP}$ holds. 
Fig.~\ref{p3D} illustrates this calculation for a specific choice of the model 
parameters (see figure caption).
%
%
%
%
The HP$\rightarrow$MP phase transition takes place if the pressure of the charge neutral HP (white line)
meets the pressure surface of the QP. This happens at point \boxed{1} in Fig.~\ref{p3D}. Up to this point
the pressure of the QP is below the pressure of the HP, making the HP the physically realized one.
Up to point \boxed{2} the phase follows the MP curve which is given by the 
Gibbs condition (\ref{GibbsCondition}). At point \boxed{2} the MP curve meets the charge neutral QP curve
(white line) and the pressure of the QP is above the pressure of the HP, making the QP the physically
realized one. The resulting chemical potentials are again shown in Fig.~\ref{munmue}.
%
%
%
Below the charge neutral HP curve and above the charge neutral
QP curve the HP is positively charged  (\mbox{$\rho_c^{HP}>0$}, $\rho_c$ denotes the 
charge density) and the QP is negatively
charged \mbox{($\rho_c^{QP}<0$)}. Therefore, the charge of hadronic matter can be neutralized
in the MP by an appropriate amount of quark matter.
Up to the critical chemical potential $\mu_n^{[1]}$ the electron chemical potential of the 
charge neutral HP is increasing. (This is mainly because electrons have
to neutralize the increasing amount of protons appearing in the HP.)
At $\mu_n^{[1]}$ the phase transition to the MP takes place and $\mu_e$ can be reduced due to
the first appearance of negatively charged quark phase droplets immersed in the hadronic phase. 
(For a detailed discussion of the geometrical structure of the MP see \cite{GlenBook}.)
For every point on the MP curve one has to calculate the volume proportion
\begin{equation} 
\chi = \frac{V_{QP}}{V_{QP}+V_{HP}}
\end{equation}
occupied by the quark phase by imposing the condition of global charge neutrality in the MP
\begin{equation} \label{globalcharge}
\chi \, \rho_c^{QP} + (1-\chi) \, \rho_c^{HP} = 0.
\end{equation}
From this, the energy density $\epsilon_{MP}$ of the MP follows as
\begin{equation}\label{epsilonQP}
\epsilon_{MP} = \chi \, \epsilon_{QP} + (1-\chi) \, \epsilon_{HP}.
\end{equation}
In the MP the volume proportion of the quark phase is monotonically increasing 
from $\chi=0$ at $\mu_n^{[1]}$
to $\chi=1$ at $\mu_n^{[2]}$ where the  transition into the pure QP takes place. 
Taking the charge neutral EOS of the HP (Sec.\,\ref{SectRMF}) for $\mu \le \mu_n^{[1]}$, Eq. 
(\ref{GibbsCondition}), (\ref{globalcharge}) and (\ref{epsilonQP}) for the MP for  
$\mu_n^{[1]}<\mu_n<\mu_n^{[2]}$ and the charge neutral EOS of the QP  (Sec.\,\ref{SectQP}) 
for $\mu_n \ge \mu_n^{[2]}$
we can construct the full EOS in the form $p=p(\epsilon)$. For simplicity we denote the complete EOS
as the {\em hybrid star} EOS.

As one realizes from Fig.~\ref{p3D}, the phase transition construction in a two-component system 
leads to a monotonically increasing pressure of the MP with increasing density \cite{Glen92}. This 
is quite different from the constant pressure in a one-component system 
(with only one independent chemical 
potential) which neutron star matter is occasionally assumed to be. 
For the existence of a MP inside the star an increasing pressure is crucial. In turn, 
such a MP allows for a softer EOS over a wide density range which would be excluded in the
one-component treatment. We will 
discuss this point in more detail in the appendix. There we show that a
simplified treatment of the phase transition with only one independent component results 
in an overestimation of the phase transition density. 

\subsection{Results for the equation of state}\label{ResultsEOS}

\begin{figure}[ht]
\centerline{\epsfig{file=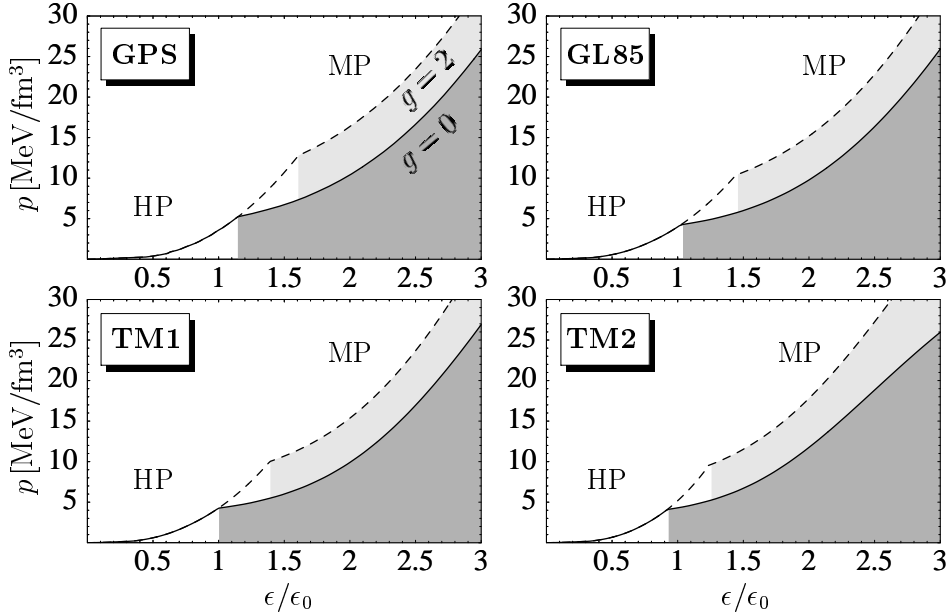,width=\figwidth}}
\caption{Low density range of the EOS for several hadronic EOS and two coupling 
constants $g$ of the quark phase. The bag constant of the quark phase is 
chosen as \mbox{$B^{1/4}=170\,$MeV}.
Shaded regions correspond to the mixed phase (MP) parts of the EOS. 
HP denotes the hadronic phase. 
\mbox{$\epsilon_0=140\,$MeV/fm$^3$}.}
\label{EOSlow}
\end{figure}

\begin{figure}[ht]
\centerline{\epsfig{file=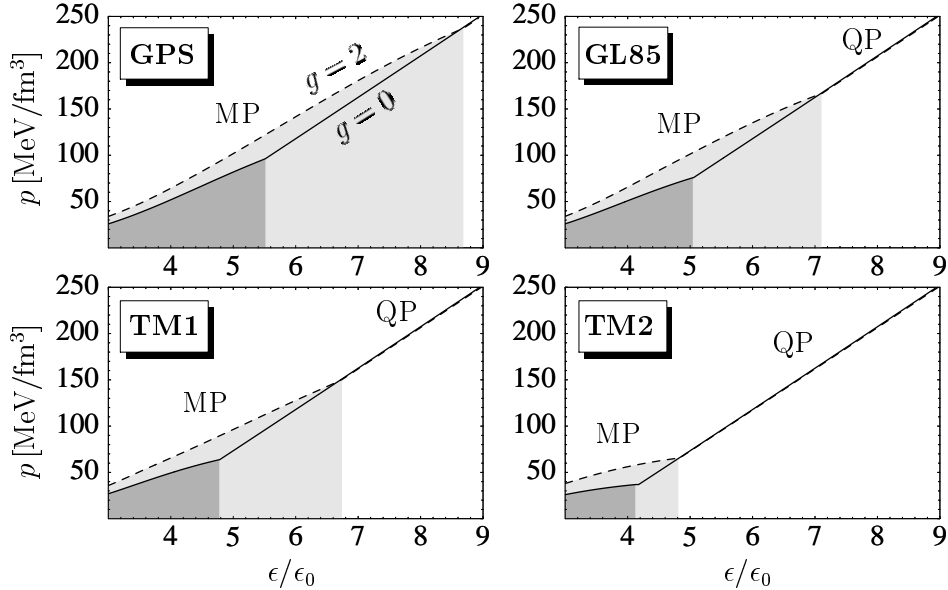,width=\figwidth}}
\setlength{\fboxsep}{2pt}
\caption{High density range of the hybrid star EOS of Fig.~\ref{EOSlow}.
The bag constant of the quark phase (QP) is chosen as $B^{1/4}=170\,$MeV.
Shaded regions correspond to the mixed phase (MP) parts of the EOS. 
$\epsilon_0=140\,$MeV/fm$^3$.}
\label{EOShigh}
\end{figure}

In this section we present the results of the phase transition construction discussed in the last
section. We study in particular the influence of different hadronic EOS and the influence of
the model parameters of the QP on the properties of the phase transition. 
Fig.~\ref{EOSlow} shows the low
density range of the hybrid star EOS in the form $p=p(\epsilon)$ for the four hadronic EOS 
and two coupling constants of the QP.
%
%
%
The bag constant is chosen to be \mbox{$B^{1/4}=170\,$MeV}.
The MP part of the EOS is shaded gray. As already noted, even in the 
MP the pressure is monotonically increasing with increasing density. 
For that reason the MP does exist in the interior of the neutron star if its central
energy density exceeds the HP$\rightarrow$MP transition density.
One can see that the HP$\rightarrow$MP transition
density depends only slightly on the choice of the hadronic EOS. This reflects the rather
small uncertainty in the hadronic EOS in the density range up to about $1.5\,\epsilon_0$
where hyperonic degrees of freedom are still not present. 
(In the HP the hyperons typically appear at $2\!-\!3\,\epsilon_0$ \cite{SchaMish96}.)
The HP$\rightarrow$MP transition density also depends only weakly 
on the influence of medium effects which shift the
transition densities to slightly higher values. This was already found in \cite{Sche98}. 
Obviously the EOS gets softer compared to the HP due to the onset of the 
phase transition. (By ``softer'' we denote a smaller pressure at a fixed
energy density $\epsilon$). In Fig.~\ref{EOShigh} we show the same 
hybrid star EOS in the high density region.
%
%
%
In contrast to the HP$\rightarrow$MP transition density we see that the MP$\rightarrow$QP transition 
density is quite
sensitive to the choice of the hadronic EOS as well as to the influence of medium effects in the QP. 
Assuming e.g.~a star with a central energy density of $\epsilon=5\,\epsilon_0$ we see that no conclusive
statement can be made about the composition of the center of the star. Neglecting medium effects ($g=0$),
TM1 and TM2 predict a QP core (QC) while GPS and GL85 predict a MP core (MC). Taking medium
effects into account ($g=2$) only TM2 predicts a QC. The strong sensitivity of the transition densities
can qualitatively be understood if we look again at Fig.~\ref{p3D}. In the high pressure region where the
MP$\rightarrow$QP transition takes place, the HP and the QP pressure surfaces are nearly parallel.
Therefore a small change in one EOS is able to produce considerably large changes in the 
MP$\rightarrow$QP transition density.
Furthermore one should note in Fig.~\ref{EOShigh} that (despite a change in the phase
transition densities) the influence of medium effects 
on the pure QP EOS in the form $p=p(\epsilon)$ is negligible. 
This was already found in \cite{Sche97}.

\begin{figure}[tt]
\centerline{\epsfig{file=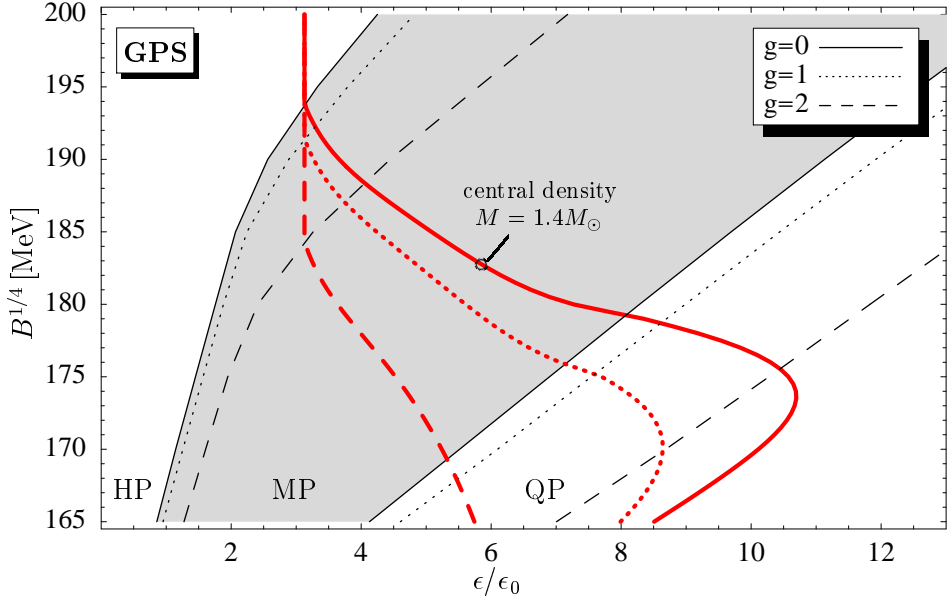,width=\figwidth}}
\caption{Phase transition densities  
for three different coupling constants $g$ (thin lines) as a function of the bag constant $B$.
The shaded region (MP) marks the density range of the MP at $g=0$. The thick lines show
the central energy density of a $1.4 M_\odot$ star. As long as the thick line is to the right
of the corresponding MP range (e.g.~the shaded region for $g=0$) 
the star possesses a pure quark phase core.
$\epsilon_0=140\,$MeV/fm$^3$.}
\label{BepsGPS}
\end{figure}

\begin{figure}[tt]
\centerline{\epsfig{file=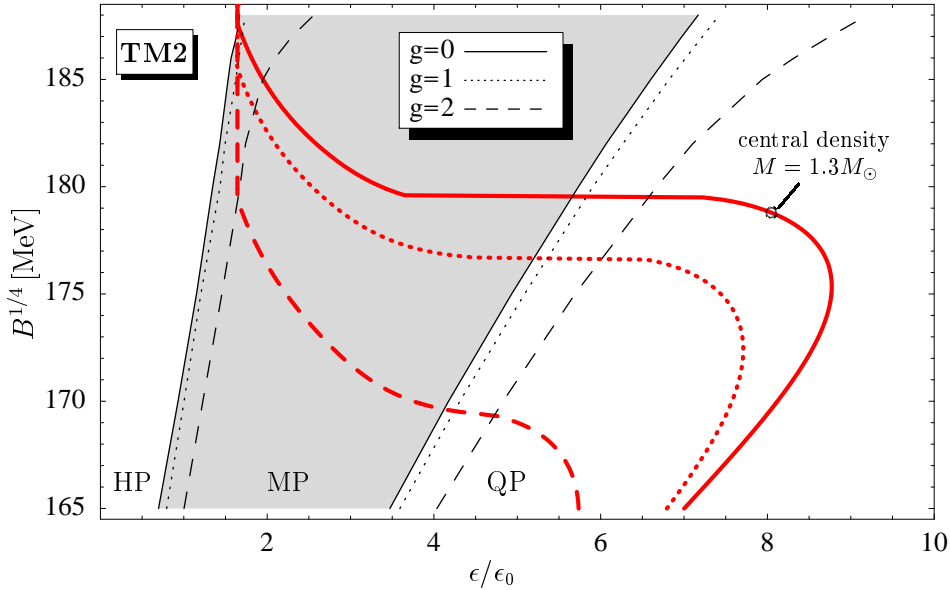,width=\figwidth}}
\caption{Phase transition densities  
for three different coupling constants $g$ (thin lines) as a function of the bag constant $B$.
The shaded region (MP) marks the density range of the MP at $g=0$. The thick lines show
the central energy density of a $1.3 M_\odot$ star. 
$\epsilon_0=140\,$MeV/fm$^3$.}
\label{BepsTM2}
\end{figure}

Up to now we have neglected the influence of the bag constant $B$ which enters in the QP EOS. 
Since $B$ enters negatively into the QP pressure in Eq. (\ref{QPepsp}) a larger $B$ will shift the pressure
surface $p_{QP}(\mu_n,\mu_e)$ in Fig.~\ref{p3D} down which moves the MP intersection curve to higher
pressures. This finally also shifts the transition densities to higher values. Fig.~\ref{BepsGPS} and 
Fig.~\ref{BepsTM2} demonstrate this behavior for different hybrid star EOS using GPS and TM2 for the HP 
(the other hybrid star EOS using TM1 and GL85 show a very similar behavior). 
%
%
%
%
%
%
In these figures the gray shaded area
corresponds to the density region of the MP for $g=0$ (i.e.~no medium effects). 
Obviously the HP$\rightarrow$MP and the MP$\rightarrow$QP transition 
densities are increasing with increasing 
$B$ and increasing $g$ where the MP$\rightarrow$QP transition density again is more sensitive.
An increase of $B^{1/4}$ by only $5\,$MeV is able to increase the
MP$\rightarrow$QP transition density by about one $\epsilon_0$. From Fig.~\ref{BepsGPS} it is
furthermore interesting to note that the HP$\rightarrow$MP transition density seems to be only slightly 
sensitive
to a change of the model parameters as long as its density is below $\epsilon\approx2\epsilon_0$ and 
strongly sensitive above (comparable to the MP$\rightarrow$QP transition density). This effect 
seems to be related to the appearance of hyperons (mainly $\Sigma^-$ and $\Lambda$) 
in the HP EOS at densities of 
$\epsilon \approx 2\!-\!3\epsilon_0$ which has already been noted in \cite{Sche98}.

It is important to note that the increase of the transition densities with increasing $B$
does not necessarily disfavor the existence of deconfined phases inside the star since
also the central energy density can be increased by changing this 
parameter. This is illustrated by the thick lines in Fig.~\ref{BepsGPS} and Fig.~\ref{BepsTM2} which
show the central energy density of a star at fixed mass as a function of the bag constant.
(The calculation of neutron star masses is discussed in the next section. In Fig.~\ref{BepsTM2}
we choose a mass of $M=1.3\,M_\odot$ instead of the ``canonical'' value of $M=1.4\,M_\odot$ 
since it turns out that for some ($B$, $g$) parameters the maximum mass
($M_{{\rm max}}$) is slightly below $1.4\,M_\odot$.) 
At least for $g=0$ and $g=1$ the central energy density 
is increasing with $B$ for low values. From these figures we can already
see which kind of phase (HP, MP or QP) is realized in the center of the star. E.g.~from Fig.~\ref{BepsGPS}
we find that at $g=0$ the central energy density (thick solid line)
of a $M=1.4 M_\odot$ star is above the 
MP$\rightarrow$QP transition density for bag constants below $B^{1/4}\approx180\,$MeV. In this
parameter range the neutron star possesses a pure QP core (QC). At higher $B$ up to 
$B^{1/4}\approx195\,$MeV the star possesses a MP core (MC). At still higher $B$ the star is purely
hadronic (HC). We will come back to a discussion of the ($B$, $g$) parameter range leading to one of 
the possibilities (QC, MC, HC) in the next section. 

In the following we first want to focus in some more detail on the interesting behavior of the 
central energy density with increasing $B$. The central energy density of a star at fixed mass 
depends on the softness
or stiffness of the underlying EOS. Soft EOS possess a larger central energy density 
as compared to stiff EOS.
In a simple way one could imagine that stars corresponding to soft EOS are compressed to smaller 
radii and therefore possess a larger central density. In our 
case (e.g.~Fig.~\ref{BepsGPS},  thick line $g=0$) two competing effects are important. By increasing $B$, 
i) the low density range of the hybrid star EOS gets stiffer due to an enlarged density range of the HP 
(which is stiffer than the MP) while ii) the  
density range of the QP gets softened directly by a larger bag constant. Effect ii) dominates 
for smaller $B$ since 
large parts of the star are made of the QP. Therefore the central energy density is increasing with $B$.
In both figures this holds up to about $B^{1/4}\approx175\,$MeV for $g=0$. We will see in the next section
that in this range the internal structure of the star is only slightly sensitive to 
a change of $B$ because both the MP$\rightarrow$QP transition density and the central energy
density are equally increasing with $B$.
The situation is reversed for bigger $B$ and the central energy density is decreasing towards
the value of the pure hadronic star. This behavior demonstrates two things. Firstly, 
the central energy density depends quite sensitively on the softness or stiffness of the EOS 
which effect the composition of neutron star matter. 
Thus it is not possible to talk about a ``typical central energy density'' of a neutron star without having
a particular EOS in mind.
Secondly, the increase of the bag constant -- which makes the QP energetically less favorable -- 
does not automatically disfavor the existence of a QP inside the star since also the central 
energy density is increased. Indeed, there is a region of the bag constant ($B^{1/4}\lapprox175\,$MeV) 
where the internal structure of the star depends only slightly on it. We will come back to this point below. 
We furthermore should note that the discontinuous fall of the 
central energy density in Fig.~\ref{BepsTM2} for $g=0$ at $B^{1/4}\approx180\,$MeV and for $g=1$ 
at slightly lower $B$ is related to the existence of ``neutron star twins'' which we will discuss in Sec.~\ref{Twins}. 

\section{Neutron star structure}\label{StarStructure}

With the evaluated hybrid star EOS presented in the last section we now turn to analyse the structure of the 
corresponding non-rotating neutron stars by solving the \mbox{Tolman-Oppenheimer-Volkoff} (TOV) equations 
\cite{OppeVolk39}. These equations describe the balance between the gravitational force and
the pressure given by the Fermi pressure of the particular EOS. 
This leads to a relation between the mass $M$ and the radius $R$ of the neutron star in 
general relativistic hydrostatic equilibrium.

\subsection{Mass-radius relations}\label{MRRelations}

%
%
\begin{figure}[tt]
\centerline{\epsfig{file=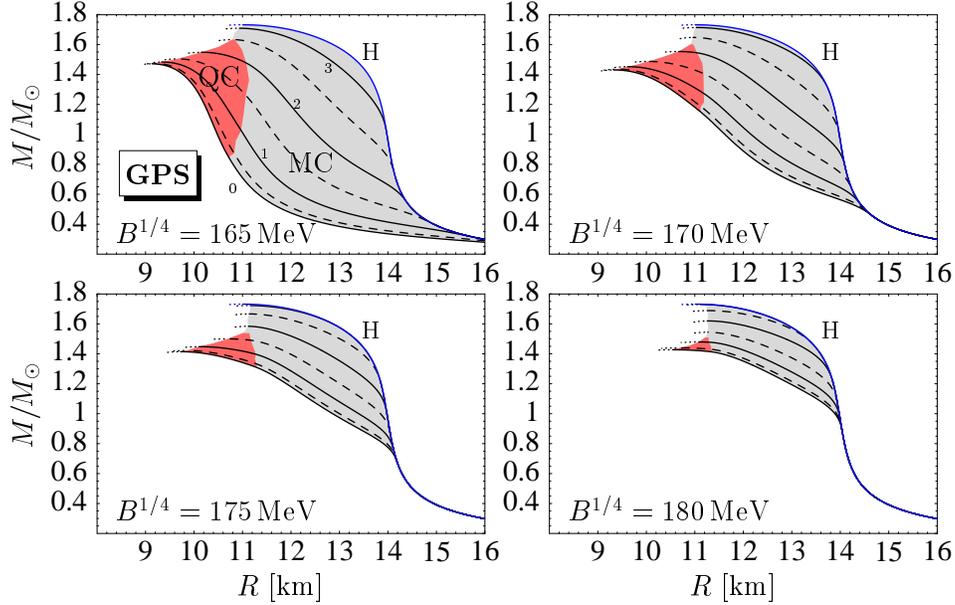,width=\figwidth}}
\caption{Mass-radius relations using the GPS EOS for the HP. The uppermost line denoted by H corresponds
to the pure hadronic solution. All other curves allow for a phase transition to the quark phase. The small
numbers denote the coupling constant $g$ employed for the QP (dashed lines denote steps of $0.5$ in $g$). 
Objects in the gray shaded MC region posses 
a mixed phase core while stars in the darker shaded QC region posses a pure quark phase core. 
Note the radius separation of about $2\!-\!4\,$km between the stars possessing a quark 
core and the pure hadronic stars of the same mass.}
\label{MRGPS}
\end{figure}
%
%
%
\begin{figure}[tt]
\centerline{\epsfig{file=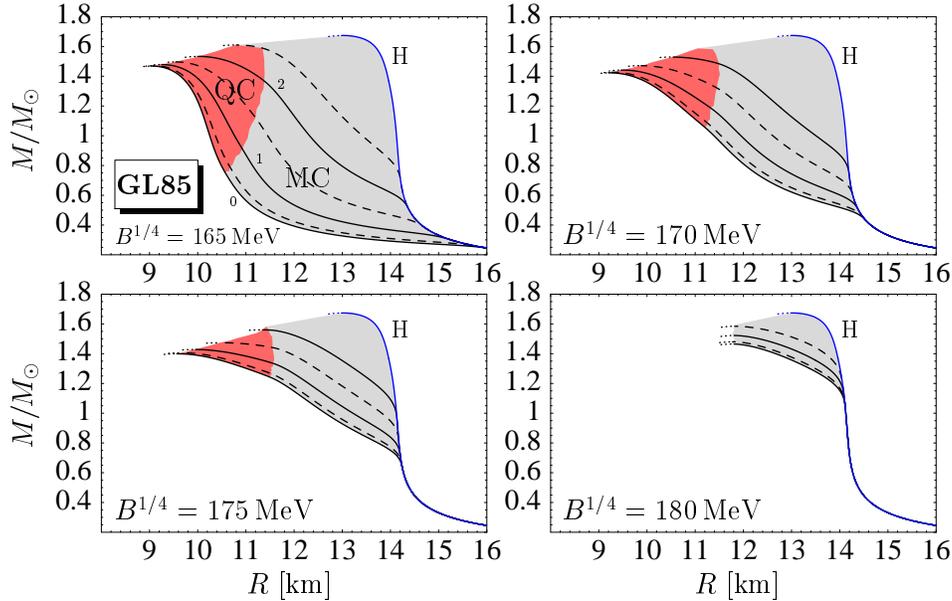,width=\figwidth}}
\caption{Mass-radius relations using the GL85 EOS for the HP. The uppermost line denoted by H corresponds
to the pure hadronic solution. All other curves allow for a phase transition to the quark phase. The small
numbers denote the coupling constant $g$ employed for the QP (dashed lines denote steps of $0.5$ in $g$). 
Stars in the gray shaded MC region posses 
a mixed phase core while stars in the darker shaded QC region posses a pure quark phase core.}
\label{MRGL85}
\end{figure}
%
Fig.~\ref{MRGPS} and Fig.~\ref{MRGL85} show the mass-radius (MR) relations for two hybrid star EOS 
(GPS and GL85) and several  
bag constants and coupling constants of the QP. The MR-relation denoted by H corresponds 
to the solution for the pure hadronic star (without any phase transition). The central energy density 
$\epsilon_c$ of the stars is increasing from large radii and small masses (down to the right) 
to the upper left edge where
the {\em critical} central energy density $\epsilon_c^{crit}$ is reached at the maximum mass and the minimal
radius of the star. From the two differently shaded regions denoted by QC (QP core) and MC (MP core)
one sees whether $\epsilon_c$ is large enough to possess a QC or a MC in the center of the star.
For small bag constants ($B^{1/4} \lapprox 180\,$MeV) and small coupling constants (lower curves) $\epsilon_c$ 
is sufficient to exceed the MP$\rightarrow$QP transition density before $\epsilon_c^{crit}$ is reached
(cf.~Fig.~\ref{EOShigh}). This leads to a pure QC inside the star. 
For a more detailed discussion of MR-relations and in particular of the influence of medium effects on the 
MR-relation see \cite{Sche98}.

It is now our major aim to see what the effect of a pure QC on the mass and the radius of a neutron star is.
Concerning the mass of the star we find that $M_{{\rm max}}$ of a pure hadronic star (H) 
is reduced by about $0.1\!-\!0.3\,M_\odot$ due to the phase transition and the corresponding softening
of the EOS. Nevertheless, both hybrid star EOS are able to explain a typical neutron star mass of about
$1.4\,M_\odot$. The most compact objects seem to have radii of about $9\!-\!10\,$km. In our model
calculations these radii are only 
reached in stars possessing a pure QC (especially for $g\approx0$). Very similar 
results are found for the two other hybrid star EOS (TM1 and TM2) which are not shown here.
If we furthermore compare the radii of stars located in the QC region with the radii of the pure hadronic stars 
(H) of the same mass we can see that the QC stars are about $2\!-\!4\,$km smaller than 
the corresponding hadronic stars. 
Hence, a neutron star is about $20\!-\!30\%$ more compact due to the 
presence of a pure QP in its center.
%
%
\begin{figure}[tt]
\centerline{\epsfig{file=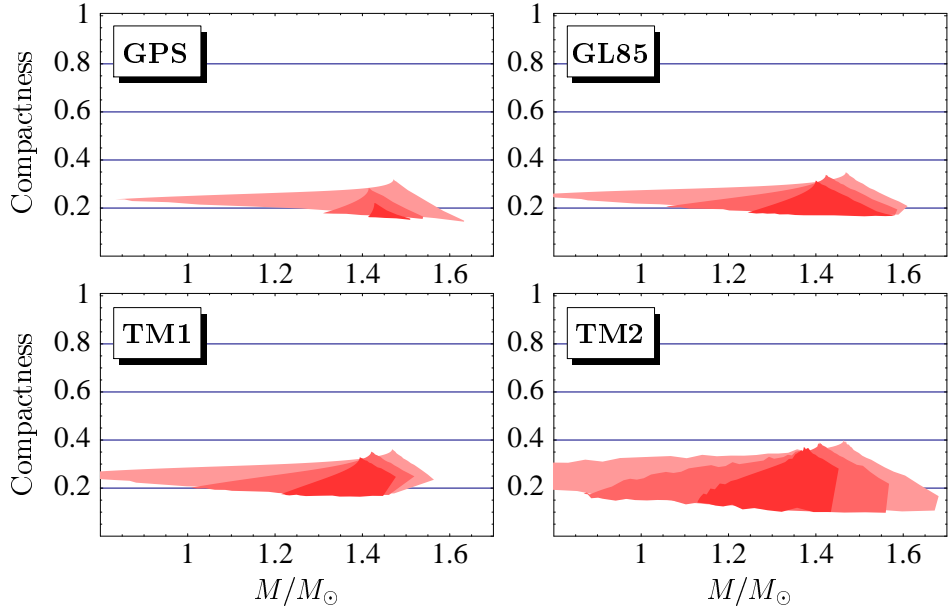,width=\figwidth}}
\caption{Compactness (\ref{defcompactness}) of neutron stars possessing a pure 
quark core with respect to the pure hadronic star of the same mass.
Obviously the presence of quark phases inside the star reduce the radius by typically $20\!-\!30\%$. 
The shaded areas stem from varying the model parameters of the quark phase 
(bag constant $B$ (shading) and coupling constant $g$) which lead to a pure quark core inside the star. 
Especially for GPS, GL85
and TM1 the compactness ratio seems to be rather insensitive to the particular hadronic EOS and to the 
model parameters of the quark phase.
} \label{Compactness}
\end{figure}
%
We define the {\em compactness} of a neutron star possessing a QC as
\begin{equation}\label{defcompactness}
\mbox{compactness(M)} \equiv \frac{R_{H}(M)-R_{QC}(M)}{R_{H}(M)},
\end{equation}
with the radius $R_{QC}$ of the QC star and the radius $R_{H}$ of the pure hadronic
star of the same mass (cf.~Figs.~\ref{MRGPS}, \ref{MRGL85}). 
The compactness therefore corresponds to the ratio to
which the radius of a pure hadronic star can be reduced due to the presence of a pure QP. 
The shaded areas in Fig.~\ref{Compactness} come from calculating the compactness for all stars located
in the QC region of the corresponding MR-relation (cf.~Figs.~\ref{MRGPS}, \ref{MRGL85}) 
for different
bag constants (which are depicted in Fig.~\ref{Compactness} by different shadings). 
It is remarkable that the compactness of $0.2\!-\!0.3$ depend only slightly on the particular HP EOS 
and on the model parameters applied in the QP. Especially an increase of the 
bag constant $B$ (darker shading) 
only narrows the mass range leading to a QC. At the same time the range of the compactness 
is only slightly altered. A reduction of the radius 
of an hadronic star by $20\!-\!30\%$ due to the presence of a pure QP seems to be a typical value. 
Only for $g\approx0$ (which in general leads to the most compact stars) 
and the TM2 hybrid star EOS a radius reduction of about $40\%$ seems possible 
(cf.~Fig.~\ref{Compactness}). 
This might be related to a rather large incompressibility $K$, a small saturation density $\rho_0$, 
and a small effective mass $m^*_N$ of the TM2 EOS compared to the other EOS (cf.~Tab.~\ref{NuclProps}).

\subsection{Internal structure}

\begin{figure}[ht]
\centerline{\epsfig{file=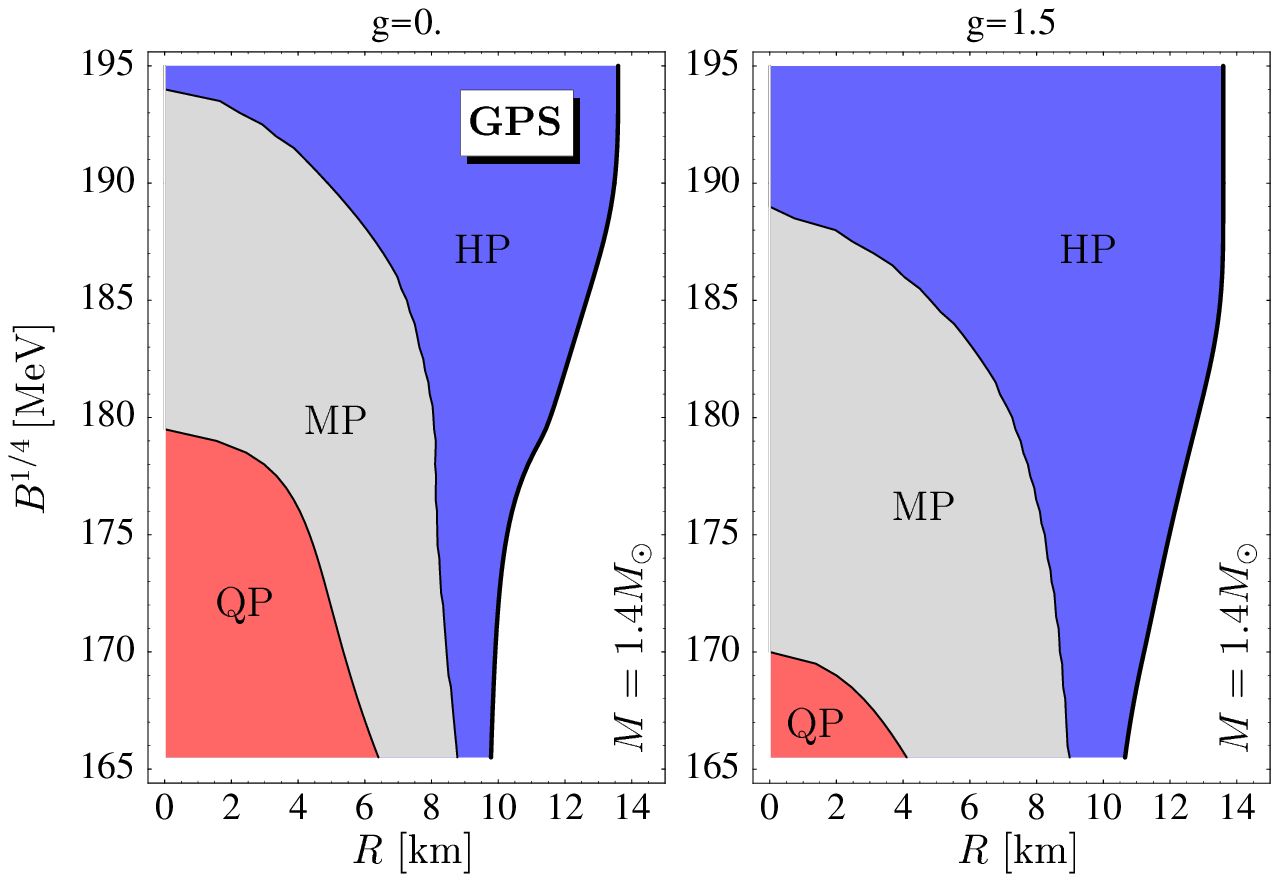,width=\figwidth}}
\caption{Internal structure of a $M=1.4 M_\odot$ neutron star as a function of the bag constant.
Left panel $g=0$ (no medium effects), right panel $g=1.5$. The hadronic EOS is GPS.}
\label{BRGPS}
\end{figure}

\begin{figure}[ht]
\centerline{\epsfig{file=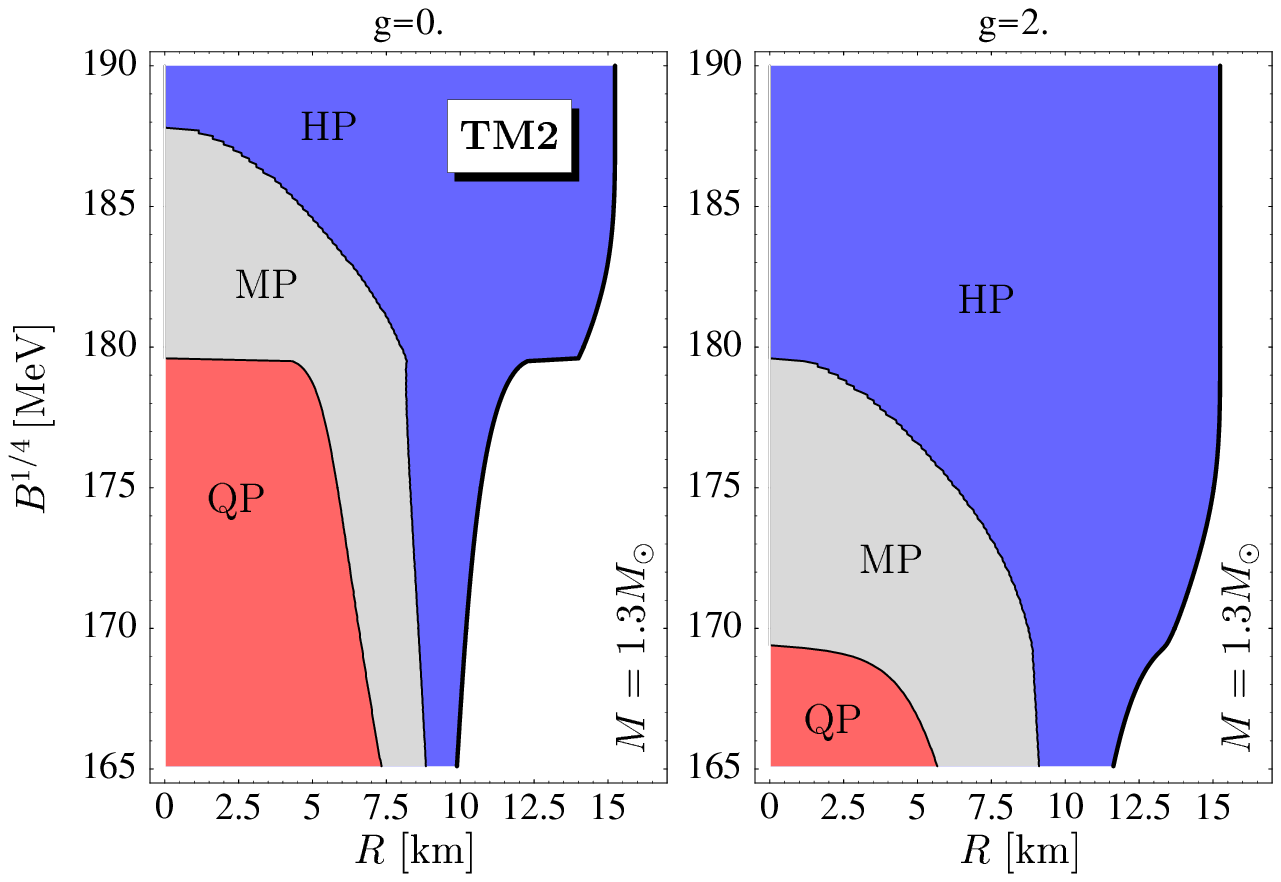,width=\figwidth}}
\caption{Internal structure of a $M=1.3 M_\odot$ neutron star as a function of the bag constant.
Left panel $g=0$ (no medium effects), right panel $g=2$. The hadronic EOS is TM2.}
\label{BRTM2}
\end{figure}

In this section we further study how the internal structure 
of the star e.g.~the radius of the quark core (QC) or 
the thickness of the MP depends on the model parameters of the QP. 
Fig.~\ref{BRGPS} and Fig.~\ref{BRTM2} show the neutron star cross sections for a range of 
bag constants $B$ and a {\em fixed} neutron star mass.
%
%
%
%
%
%
The respective left panels show the results without medium effects ($g=0$) while the right panels 
include medium effects in the QP.
The three differently shaded regions QP, MP and HP again correspond to the parts of the star made of 
quark phase, mixed phase and hadronic phase, respectively. The thick line to the right of the HP
marks the surface of the star. It is obvious that basically only the presence of a pure QP is able to reduce 
the radius of the star significantly.
For $g=0$ (left panel) we can see in both figures that only for $B^{1/4} \lapprox 180\,$MeV
a pure QC can exist. For $B^{1/4} \lapprox 175\,$MeV the radius of the QC can be as large as $4\!-\!7\,$km
at an overall radius of about $10\,$km.
For these values of $B$ we can see that the internal structure 
of the star (especially the overall radius) only slightly 
depends on a change of $B$. This behavior can be traced back 
to the findings of Sec.~\ref{ResultsEOS}. There 
we found that despite of an increase of $B$ from $B^{1/4}=165\,$MeV to $175\,$MeV 
(which increases the MP$\rightarrow$QP
transition density in the GPS case by about $3\,\epsilon_0$) 
a pure QP is not excluded due to a similar increase of the central energy density $\epsilon_c$ 
(cf.~Figs.~\ref{BepsGPS} and \ref{BepsTM2}). 

On the right panels we can see how the influence of medium
effects change the internal structure of the star. Since the HP$\rightarrow$MP 
and MP$\rightarrow$QP transition 
densities at fixed $B$ increase with increasing $g$ (cf.~Fig.~\ref{BepsGPS} and \ref{BepsTM2}) 
medium effects are able to transform a star with MC into an pure hadronic star (e.g.~Fig.~\ref{BRGPS}, 
$B^{1/4}=190\,$MeV) or a star with QC to one with a MC (e.g.~Fig.~\ref{BRGPS}, $B^{1/4}=175\,$MeV).
Only a lower bag constant can compensate for this effect to a certain degree. 
The solutions for the TM1 and GL85
hybrid star EOS (which are not shown here) lead to very similar results.

\begin{figure}[tt]
\centerline{\epsfig{file=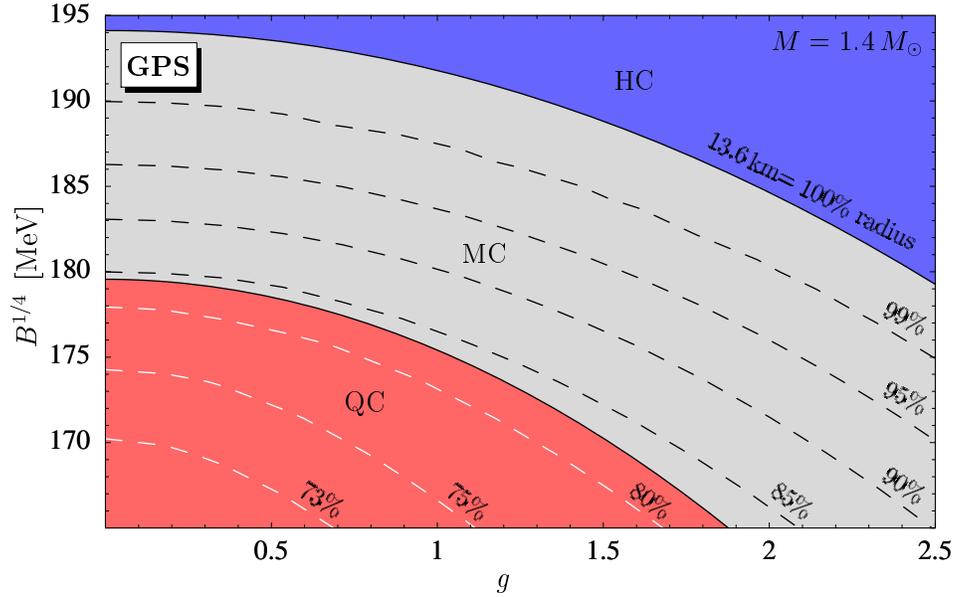,width=\figwidth}}
\caption{Parameter range (bag constant and coupling constant) 
of the internal structure of a $M=1.4 M_\odot$ neutron star. Parameters in the QC region lead to
neutron stars with pure quark core while MC or HC corresponds to a mixed core or a pure hadronic star, 
respectively. The dashed lines are lines of constant radius given as the percentage of the radius of 
the hadronic star. The hadronic EOS is GPS.}
\label{BgGPS}
\end{figure}

\begin{figure}[tt]
\centerline{\epsfig{file=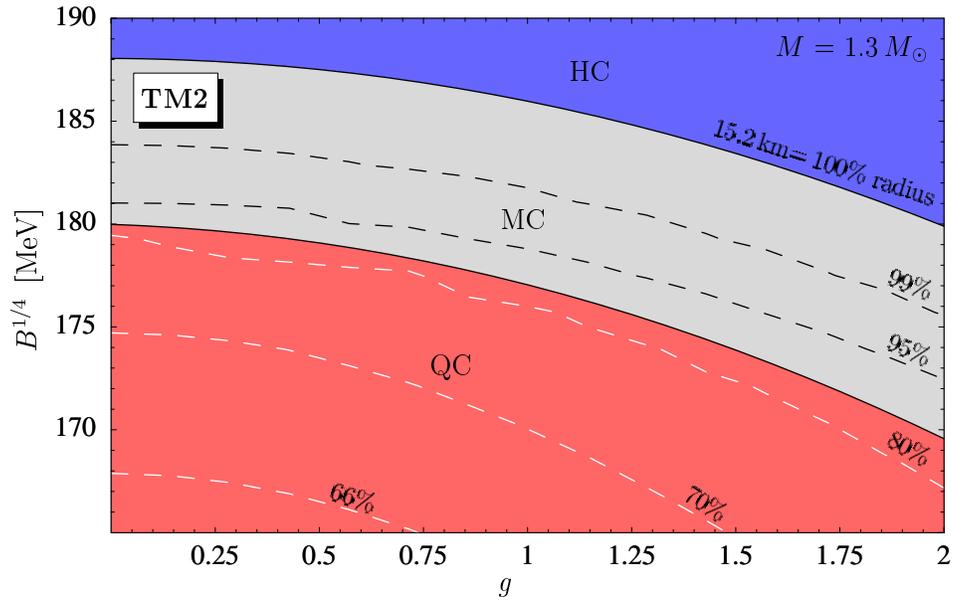,width=\figwidth}}
\caption{Same as Fig.~\ref{BgGPS} but for a neutron star mass of $M=1.3 M_\odot$. The hadronic EOS is TM2.}
\label{BgTM2}
\end{figure}

Finally, we illustrate the complete ($B$, $g$) parameter range in Fig.~\ref{BgGPS} and \ref{BgTM2} 
to see which values lead to which kind of internal structure (QC, MC or HC (pure hadronic star)).
%
%
%
%
%
%
Again we can see that - if we want to keep 
the internal structure of the star nearly fixed - the influence of medium effects can only be 
compensated by lowering the bag constant. The dashed lines show
the neutron star radii in terms of the radius of the pure hadronic star. 
These lines correspond to lines of constant compactness 
(\ref{defcompactness}). In both figures the borderline between the QC and the MC region 
appears at about $80\%$ (corresponding to a compactness of $0.2$).
This demonstrates once more that - almost independently of the model 
parameters of the QP - a neutron star with a MC can at most be about $20\%$ 
more compact than a hadronic star of the same mass while a
star with a QC  is typically $20\!-\!30\%$ more compact than a hadronic star.

\clearpage

\section{Phase transitions and the third family} \label{Twins}

\subsection{Gerlachs criterion} \label{Gerlach}

\begin{figure}[ht]
\centerline{\epsfig{file=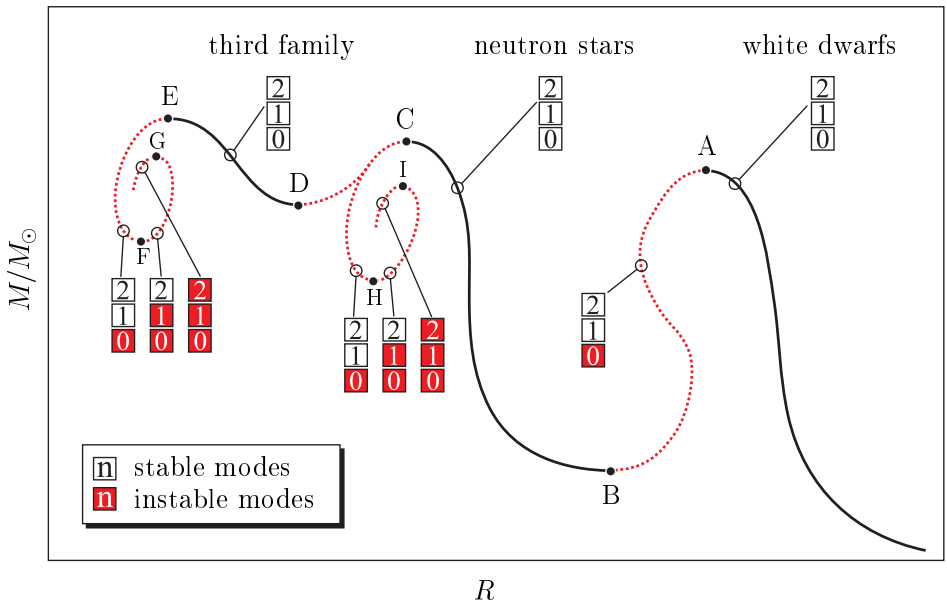,width=\figwidth}}
\caption{Schematic mass-radius relation showing three stable families of compact stars.
The letters A, B, \dots, I refer to critical points (turning points) where a vibrational mode changes
stability \cite{ShapiroBook,WheelerBook}. The stability (solid lines) or instability (dotted lines) 
of the three lowest-lying modes 
($n=0,1,2$) is depicted by the numbers. Higher modes are stable. See text for more details.}
\label{TwinsFamily}
\end{figure}

Almost 30 years ago, it was found by Gerlach \cite{Gerl68,GerlachPhD} that 
a {\em third family} of stable equilibrium
configurations of compact stars are not forbidden by general relativity. 
Such a third family could therefore in principle
exist besides the two known families of white dwarfs and neutron stars. (See Fig.~\ref{TwinsFamily} for a 
schematic view of the three families in a MR-radius relation. The criteria for stability and 
instability of the shown ranges will be discussed in Sec.\,\ref{Stability}.).
%
%
%
Gerlach found that a {\em necessary} condition for the theoretical possibility 
of this family is a sufficiently large
discontinuity in the speed of sound ($c_s^2=dp/d\epsilon$) of the corresponding 
EOS\footnote{This was proved by means of an inversion of the Tolman-Oppenheimer-Volkoff 
equations. In this context see also the work by Lindblom \cite{Lind92}.}.
As the density is increased, the speed of sound has to increase abruptly at a density exceeding the  
central energy density of the neutron star of limiting mass (point C in Fig.~\ref{TwinsFamily}). 
Rather smooth EOS (as comprehensively studied by Wheeler et al.~\cite{WheelerBook}) does not
possess the possibility of a third family.
Thus a verification of the existence of a third family in nature is
equivalent to the verification of a phase transition in the EOS of matter at large densities.

\subsection{Quark phases and the third family} \label{QuarkThird}

\begin{figure}[ht]
\centerline{\epsfig{file=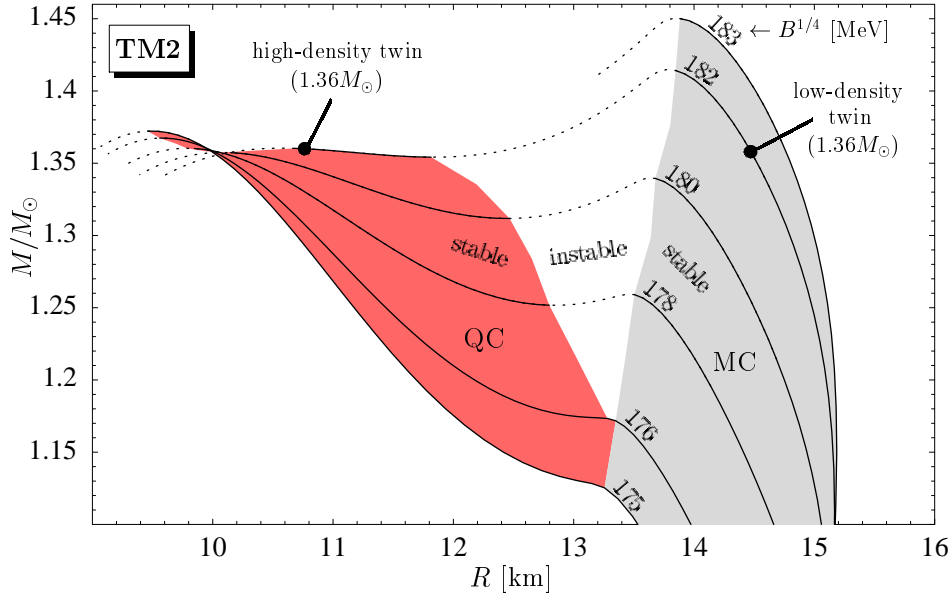,width=\figwidth}}
\caption{Mass-radius relation using the TM2 EOS for the HP for different bag constants $B$ and $g=0$.
For this EOS a third family only exists in the narrow parameter range of 
$176\,$MeV  $< B^{1/4} \le 182\,$MeV. Obviously the stars of the third family possess 
approximately the same minimal radii as an ``ordinary'' hybrid star with pure QC as 
given by $B^{1/4}=175\,$MeV.} \label{TwinsMR}
\end{figure}

\begin{figure}[ht]
\centerline{\epsfig{file=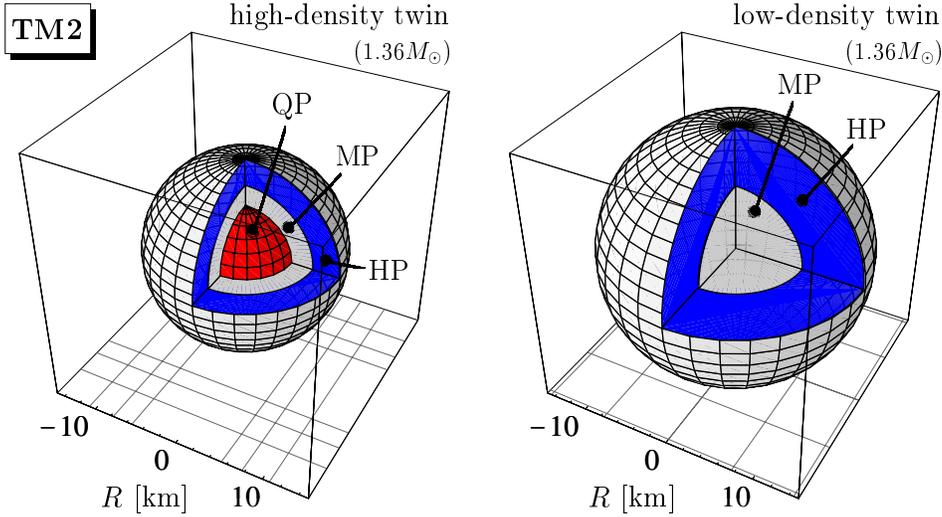,width=\figwidth}}
\caption{Schematic gross structure of the non identical neutron star twins marked in Fig.~\ref{TwinsMR}.
Both stars have $M=1.36 M_\odot$ and belong the same EOS 
(TM2 hybrid star EOS, $B^{1/4}=182\,$MeV, $g=0$).}
\label{Twins3D}
\end{figure}

From Gerlachs criterion it remains however unclear i) which physical mechanism could produce a discontinuity 
(a phase transition) sufficient to support 
the stability of a third family and ii) whether a formation process exists in nature 
to physically realize these objects.  
Concerning i) it was recently found by Glendenning and Kettner \cite{GlenKett98}
that certain EOS that describe a first-order phase transition to deconfined quark matter (like the EOS
also considered in this work) could indeed lead to a third family.
Due to partially overlapping mass regions of the neutron star branch and the branch of the third family it is
possible that non-identical stars of the same mass can exist. Such pairs are refered 
to as ``neutron star twins'' \cite{GlenKett98}. 
In such kind of EOS the necessary discontinuity in the speed of sound is produced 
by the MP$\rightarrow$QP transition
of the deconfinement phase transition. 
As can be seen in our model from Fig.~\ref{EOShigh}, especially 
the TM2 hybrid star EOS possesses a large discontinuity 
due to a rather low speed of sound at the high density end of the MP. In a particular parameter range
of the bag constant we will see that the TM2 hybrid star EOS 
indeed allows for the existence of a third family\footnote{For the other hybrid 
star EOS we have checked more than 200 MR-relations following from
different combinations of $B$ and $g$ without finding any further solutions including a third family. 
Since the parameter range for a third family seems to be quite small we might however ignored some solutions.}. 
Fig.~\ref{TwinsMR} shows MR-relations using the TM2 hybrid star EOS. They
are shown for six different bag constants $B^{1/4}$ from $175$ to $183\,$MeV. 
%
%
We furthermore assume $g=0$ (no medium effects in the QP). The solutions for
$B^{1/4}=175\,$MeV and  $B^{1/4}=183\,$MeV do not lead to a third family and thus are
similar to QC and MC solutions found in Figs.~\ref{MRGPS} and \ref{MRGL85}. 
Increasing the central density beyond the critical point does not restore stability
(corresponding to range C-H-I in Fig.~\ref{TwinsFamily}). This is not the case 
for bag constants in the small intermediate
range $176\,$MeV  $< B^{1/4} \le 182\,$MeV. There the mass-radius relations are splitted
by an instable (dotted) range which corresponds to C-D in Fig.~\ref{TwinsFamily}.
This enables the possibility of neutron star twins as shown e.g.~for the
low-density and high-density twin of mass $M=1.36 M_\odot$. While the low-density twin
(on the neutron star branch)
has a MC the high-density twin of the third family is more compact and possesses a QC. 
This is a general characteristic of neutron star twins \cite{GlenKett98}. The neutron star branch
terminates in the MP owing to a small adiabatic index $\Gamma=(\epsilon+p)/p \cdot dp/d\epsilon$
(cf.~Fig.~\ref{EOShigh}) \cite{GlenKett98}. Thus, all stars on the neutron star branch are MC stars 
(or HC stars at lower $\epsilon_c$). 
Only the larger adiabatic index 
of the QP can restore stability and therefore enables the QC stars of the third family.
The internal structure of the neutron star twins with $M=1.36 M_\odot$ 
is schematically shown in Fig.~\ref{Twins3D}.
%
%
%

In this section we have seen, that from the deconfinement phase transition described in our model a
sufficiently large discontinuity in the speed of sound arises to allow a third family of compact stars. 
This possibility  was first realized by Glendenning and Kettner \cite{GlenKett98}. 
Despite the open questions concerning the existence of a formation process,
we now want to ask for the theoretical possibility of identifying a third family  
by means of mass and radius measurements of neutron stars. 
To do so we have first to discuss the stability criteria which ultimately
separate the third family branch of a MR-relation (range D-E in Fig.~\ref{TwinsFamily}) from the 
neutron star branch (range B-C) in the same way as neutron stars are separated from white dwarfs.

\subsection{Stability criteria} \label{Stability}

While hydrostatic stability of stellar configurations is assured by means of the 
Tolman-Oppenheimer-Volkoff equations \cite{OppeVolk39} a prove of dynamical stability
requires an additional analysis of the radial vibration modes (acoustical modes) of
the star. Due to dynamical instabilities no stable stars can have central densities
in the range of about $10^9-10^{14}\,$g/cm$^3$. This separates the family of
the white dwarfs from the neutron star family.
A configuration is stable against small perturbations around hydrostatic equilibrium if and only
if the squared frequency eigenvalue $\omega_0^2$ of the fundamental vibrational mode is positive  
\cite{ShapiroBook,WheelerBook}. 
Dynamical stable and instable parts of a schematic MR-relation
are shown in Fig.~\ref{TwinsFamily}.
The letters A, B, \dots, I refer to critical points (turning points) where 
$dM/d\epsilon_{c}=0$ holds. The {\em central} energy density $\epsilon_{c}$ 
is rising from right to left along the curve A-C-I and A-C-G respectively.
Reaching a critical point one radial vibrational mode (characterized by its number of nodes $n$) 
has to change stability. This means that the corresponding squared frequency 
eigenvalue $\omega_n^2$ changes its sign \cite{ShapiroBook,WheelerBook}. 
(A stable mode requires a positive squared frequency $\omega_n^2>0$).
At critical points where $dR/d\epsilon_c<0$ holds (all points except H and F) 
an even mode changes stability while for $dR/d\epsilon_c>0$ (points H and F) an odd mode changes stability
\cite{ShapiroBook,WheelerBook}.
Starting from a stable neutron star located in the range B-C (all $\omega_n^2>0$)
we reach the critical point C by increasing $\epsilon_c$. At point C an even mode changes 
stability due to $dR/d\epsilon_c<0$ and therefore gets instable. Since 
\begin{equation} \label{omegas}
  \omega_0^2<\omega_1^2<\omega_2^2<\cdots<\omega_n^2
\end{equation}
holds  \cite{ShapiroBook,WheelerBook} this only can
be the fundamental $n=0$ mode. A perturbation would cause such instable star to explode or collapse to a 
black hole. 
The only way to recover stability and therefore to make a third family possible at larger density 
is a further change of stability of the even $n=0$ mode at point D. (Note that point H can only change 
an odd mode). 
Due to (\ref{omegas}) no even mode expect $n=0$ can change stability at this critical point 
since all higher modes are bound by $\omega_1^2$ which is still positive. For that reason the
region D-E corresponding to the third family is again stable against radial vibrations.
Without a sufficiently large discontinuity in the speed of sound (i.e.~for any reasonably smooth EOS) the
mass-radius relation follows the curve C-H-I \cite{WheelerBook,KettWebe95}. 
Increasing the central density then leads to
a successive excitation of more and more unstable modes. (Only modes up to $n=2$ are shown in 
Fig.~\ref{TwinsFamily}).

As a important consequence of the stability analysis above, the mass and radius of
point C has to be larger than the mass and radius of point D. 
Therefore it is possible that stars of the neutron star branch 
can in principle exist which are larger and heavier than stars of the third family. 

\subsection{The third family as a possible signature for a phase transition} \label{Signature}

\begin{figure}[t]
\centerline{\epsfig{file=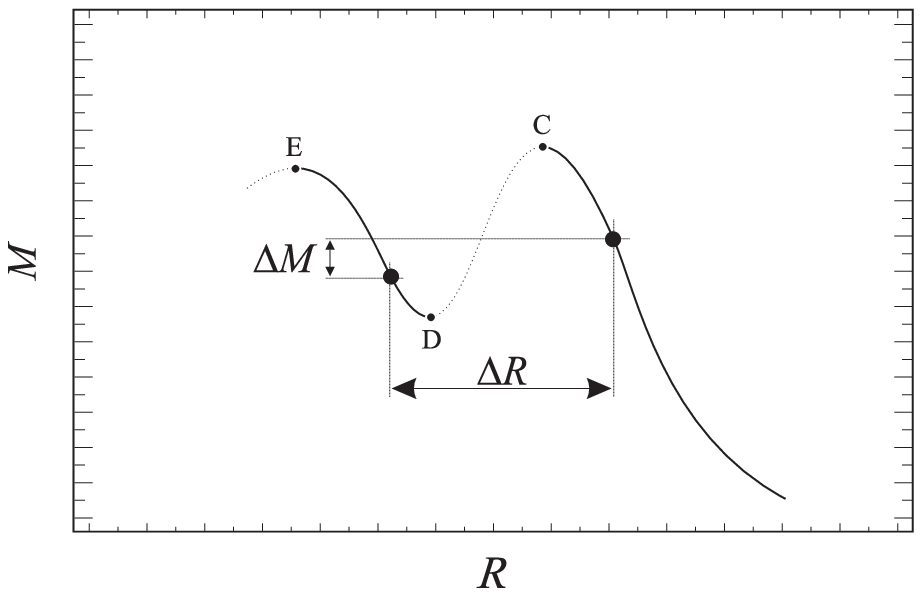,width=8.5cm}}
\caption{Schematic mass-radius relation showing a "rising twin" which can only exist if
a third family of compact stars exists ($\Delta M \ll \Delta R$).}
\label{RisingTwin}
\end{figure}

Within our model we see from Fig.~\ref{TwinsMR} that the masses and radii of the third family 
are comparable to the ones of the neighbouring $B^{1/4}=175\,$MeV MR-relation which does not possess 
a third family. 
Moreover, for a radius and mass of $R \approx 10\,$km and $M \approx 1.35 M_\odot$ we can see
the interesting behavior that all branches of the third family intersect with the branch 
($B^{1/4}=175\,$MeV) of the neutron star family in just one point.
Only the existence of such an example (even though found in a particular model)
illustrates that the measurement of the mass and the radius of only one
star (even if exact) could in general not provide the information to decide whether this star is 
a neutron star or an object of the third family. But how much points on a MR-relation do we in principle
have to know (to measure) to decide whether or not a third family exists?
The answer is {\em two}\footnote{For this it is important to note that the two masses and radii
refer to two stars located on the {\em same} MR-relation. Therefore we can e.g.~not refer to two stars
with largely different temperature, magnetic field or rotational period.}.
Comparing two arbritay points on a neutron star MR-relation 
(e.g.~$B^{1/4}=175\,$MeV) we see that the heavier star always 
possesses the smaller radius. 
But this must not hold for MR-relations including a third family as discussed in the previous section. 
Then it is possible that the heavier star possesses the larger radius. Such pair of stars always exists if
a third family exists. For that it is necessary that the smaller star is located on the third family 
branch while the larger star
is on the neutron star branch. This is schematically shown in Fig.~\ref{RisingTwin} where we have used
the same notation as in Fig.~\ref{TwinsFamily}.
%
%
%
We will refer to such pair as a {\em rising twin}. The idea of a rising twin as a signature 
for a third family is therefore based on the decrease of the mass with increasing radius
of all MR-relations without a third family.
This holds for almost all {\em gravitational bound} stars in which an increase of mass 
leads to a decrease of the radius due to the increasing gravitational attraction 
(see also the Figs.~\ref{MRGPS}, \ref{MRGL85}). 
An exception of this is given
by {\em self-bound} stars like the hypothetical strange stars which show the behavior of increasing
radius with increasing mass (for small masses $M\propto R^3$ holds) \cite{GlenBook}.
Also in some hadronic EOS (see e.g.~\cite{Akma98}) mass ranges of slightly increasing radius 
with increasing mass can exist.
However, we can exclude these exceptions by additionally require that 
$\Delta M \ll \Delta R$ ($G=c=1$) has to hold for a rising twin. 
While in the case of a third family even $\Delta M=0$ is 
allowed (the neutron star twin) this is not possible for the exceptions for which typically 
$\Delta M \gg \Delta R$ holds in the mass range of typical neutron stars. 
In the framework of our model (see Fig.~\ref{TwinsMR})
the corresponding radius differences can grow as large as $\Delta R\approx 3\,$km 
at mass differences smaller than $\Delta M\approx 0.05\,M_\odot\approx 0.07\,$km ($M_\odot \approx 1.48\,$km). 
For all possible rising twins of Fig.~\ref{TwinsMR} $\Delta M$ is more than one order of magnitude smaller
than $\Delta R$. The requirement of small mass differences is furthermore not in contradiction
to the experimentally known masses of neutron stars which are (at least for the classes of observed stars)
in a remarkably narrow range of $M=1.35\pm0.04 M_\odot$ \cite{Thor}. In particular the mass differences
of the two stars in some double neutron star binaries seems to be quite small\footnote{
For example the mass difference of the Hulse-Taylor pulsar 
PSR B1913+16 and its companion is $\Delta M\approx 0.05\,M_\odot$ while the difference of PSR B1534+12 
and its companion is even of the order of only $\Delta M\approx 10^{-3}\,M_\odot$ \cite{Thor}. 
Of course the two stars of a rising twin must not necessarily build a binary system.}.

In applying the stability considerations of Sec.~\ref{Stability}
we conclude that the detection of a rising twin (by the measurement 
of masses and radii of two neutron stars)
can prove the existence of a third family of compact stars. 
Gerlachs criterion (Sec.~\ref{Gerlach}) shows that 
this is equivalent to the existence of a phase transition 
in the EOS at finite density. The deconfinement phase
transition (as discussed in this work) is one possible scenario 
that might explain the hypothetical 
third family of compact stars (Sec.~\ref{QuarkThird}).

\section{Summary and conclusion}\label{Summary}

We have studied the properties of non-rotating cold neutron stars including the
possibility of a phase transition to a deconfined quark phase (QP). To describe the confined 
hadronic phase (HP)  of the star we have applied several hadronic EOS in 
the framework of the relativistic mean-field 
model (cf.~Sec.~\ref{SectRMF}, \cite{SchaMish96,Ghosh95}). 
The QP was modeled by an extended MIT bag model (effective mass bag model) which
includes medium effects by means of effective medium dependent quark masses 
(cf.~Sec.~\ref{SectQP}, \cite{Sche97,Sche98}). The influence of 
medium effects was parametrized by the strong
coupling constant $g$. The construction of a first-order phase transition from 
confined to deconfined matter was performed by taking into account two 
independent chemical potentials (components) of neutron star matter (cf.~Sec.~\ref{Phasetransition}, 
Fig.~\ref{p3D}). This construction allows for the existence of a mixed phase (MP) of quark and hadronic 
matter over a finite range inside
the star. We have systematically studied the influence of the model parameters
of the QP (bag constant $B$, coupling constant $g$) on the 
EOS (see Figs.~\ref{EOSlow}, \ref{EOShigh}),
the phase transition densities (see Figs.~\ref{BepsGPS}, \ref{BepsTM2}), 
the mass-radius relations of the
stars (see Figs.~\ref{MRGPS}, \ref{MRGL85}) and the corresponding internal structure 
(see Figs.~\ref{BRGPS}, \ref{BRTM2}).  
This analysis provides us with the information which parameters lead to neutron stars with 
pure quark phase core (QC), mixed phase core 
(MC) or with no phase transition at all (HC). We found that - 
almost independent of the model parameters of the quark phase and the applied hadronic EOS - 
the radius of neutron stars possessing a MC can at most be about $20\%$ more compact than a 
pure hadronic star of the same mass.  
A neutron star possessing a MC would therefore -- at least from its radius --
be hardly distinguishable from a pure hadronic star.
Stars with a QC, however, are typically $20\!-\!30\%$ more compact 
than a hadronic star. This is depicted in Fig.~\ref{Compactness} and Figs.~\ref{BgGPS}, \ref{BgTM2}.
The limitation to the upper value of $30\%$ comes from our reasonable  requirement
that the deconfinement phase transition density should not appear below normal nuclear matter density.
This approximately corresponds to $B^{1/4}\gapprox165\,$MeV (c.f.~Figs~\ref{BepsGPS}, \ref{BepsTM2}).
Within our model, the minimal radius of a neutron star is about $9-10\,$km at a corresponding mass 
of $1.4-1.5\,M_\odot$. Such small radii of $9\!-\!10\,$km are only reached in neutron stars 
with a pure QC (as large as $4\!-\!7\,$km) and could not be explained using the pure hadronic models
applied here.
On the other hand, our models fail to explain radii of $R<9\,$km.
Although some radius estimates of neutron stars suggest quite small values 
\cite{GoldShea99, Walt9697, HabeTita95, PavlZavl97,Li95,Reyn97}, 
the experimental limits (e.g.~of distance measurements) and the uncertainties in the 
interpretation of the available data (e.g.~by blackbody fits) 
do currently not allow for a definite confirmation.
Nevertheless, a future experimental confirmation of extreme small radii would offer a
unique possibility to place new stringent constraints on the EOS at high densities.
It is therefore necessary to study the theoretical limits of the compactness of neutron stars 
with inclusion of different softening mechanisms to disentangle the various scenarios in view 
of more restrictive future radius estimates. 

In Sec.~\ref{Twins} we have discussed the theoretical possibility of a third family of compact stars
which might exist besides the two known families of neutron stars and white dwarfs 
(c.f.~Fig.~\ref{TwinsFamily}). 
Within our model,
we have shown that a deconfinement phase transition can explain the existence of such third family
(c.f.~Fig.~\ref{TwinsMR}). 
Compared to stars of the neutron star family, the stars of the third family can have similar masses 
while their radii can be up to about $3\,$km smaller.
Without refering to our particular model for the EOS we argue that the availability of mass and radius
measurements of only two compact stars can reveal the existence of a third family if the larger star
is the (slightly) more massive one. We refer to such pair of stars as a 
{\em rising twin} (c.f.~Fig.~\ref{RisingTwin}).
By means of Gerlachs criterion (Sec.~\ref{Gerlach}) the existence of a third family is equivalent  
to the existence of a phase transition in the EOS at large densities 
(which not necessarily need to be the deconfinement phase transition).
If reliable mass and radius measurements would be available, the detection of a rising twin 
could therefore serve as a novel signature for a phase transition inside neutron stars.

\medskip

{\bf Acknowledgments:} 
The authors thank S.~Leupold for helpful discussions 
and for reading the manuscript. We thank P.K.~Sahu for providing us with the GPS EOS.
K.S. acknowledges the correspondence with U.H.~Gerlach concerning the third family. 


\begin{appendix}
\section*{Appendix}

In this appendix we intend to show that a simplified treatment of the phase transition construction
like the one which appears in a one-component system 
(instead of a two-component system which neutron star
matter in fact is) leads to an overestimation of the phase transition density. 
This is worth to point out since within a simplified treatment which occassionaly is
utilized for the construction of the phase transition one might exclude the possibility 
of quark phases inside neutron stars \cite{Glen92}.
In a correct two component treatment, however, the onset of a mixed phase (MP) or a 
quark phase (QP) might already be occured
inside the star. 
Compared to the simplified treatment this results in more compact neutron star configurations
due to a larger density range occupied by the soft MP or QP EOS.

The two-component system of neutron star matter in weak equilibrium can be reduced to a one-component 
one by requiring charge neutrality of the hadronic phase and the quark phase 
independently (i.e.~locally) and not globally in the MP. 
As already discussed in Sec.~\ref{SectQP}, this reduces the number of independent chemical potentials 
(components) from two (e.g.~$\mu_n$, $\mu_e$) to one (e.g.~$\mu_n$).
(In the following one argument ($\mu_n$) denotes 
charge neutral properties while two arguments ($\mu_n, \mu_e$) denotes charged ones.) 
The condition of local charge neutrality is of course unphysical in the sense that charge neutrality
in the MP can also be achieved by means of the weaker (global) condition Eq.\,(\ref{globalcharge}). 
Furthermore such a construction is thermodynamically incorrect if applied to neutron star matter. 
Whereas the Gibbs condition 
\begin{equation}\label{GibbsConditionSingle}
p_{HP}(\mu_n) = p_{QP}(\mu_n),
\end{equation}
of a one-component system does ensure the
mechanical equilibrium and the chemical equilibrium of neutrons (and therefore of all uncharged
particles)  one finds that the chemical equilibrium of the second component ($\mu_e$) 
is in general not fulfilled in the MP i.e. (cf.~Fig.~\ref{munmue})
\begin{equation}
\mu_e^{HP}(\mu_n) \not= \mu_e^{QP}(\mu_n).
\end{equation}
Consequently, all charged particles of neutron star matter do not fulfill 
the condition of chemical equilibrium.
The Gibbs condition (\ref{GibbsConditionSingle}) leads to the familiar first-order phase transition 
with a constant pressure MP. 
Since we know from the equations of hydrostatic equilibrium - the 
Tolman-Oppenheimer-Volkoff equations \cite{OppeVolk39} - that the pressure has to increase
if we go deeper into the star, a constant pressure MP is strictly excluded from the star. That is not the case
for a two-component
phase transition calculation, since the pressure is increasing with density even in the MP 
(see e.g.~Fig.~\ref{p3D}).

To show that the one-component treatment leads to an overestimation 
of the deconfinement phase transition density
we go back to Fig.~\ref{munmue} where $\mu_n^{[1]}$ marks the critical chemical potential necessary for
a HP$\rightarrow$MP phase transition of the two-component system.
All we have to show is that at $\mu_n^{[1]}$ the one-component system is still in its HP.
In other words 
\begin{equation}\label{toshow}
p_{HP}(\mu_n^{[1]})>p_{QP}(\mu_n^{[1]})
\end{equation}
must hold since at a fixed chemical potential the phase with
the larger pressure is the physically realized one. The point 
{\setlength{\fboxsep}{1pt}\fbox{1}} in Fig.~\ref{p3D}  and Fig.~\ref{munmue} is defined by the phase
equilibrium between the charge neutral HP and the charged QP, which reads
\begin{equation}\label{point1}
p_{HP}(\mu_n^{[1]})=p_{QP}(\mu_n^{[1]},\mu_e^{[1]}). \quad\mbox{(point \boxed{1} in Fig.~\ref{p3D})}
\end{equation}
Now we know that at fixed neutron chemical potential the charge density of the QP 
is given (in units of $|e|$) by
\begin{equation}\label{chargedens}
\rho_c^{QP}(\mu_n,\mu_e)=-\frac{\partial p_{QP}(\mu_n,\mu_e)}{\partial \mu_e}.
\end{equation}
Since the QP is negatively charged at point \boxed{1} (and everywhere above the charge neutral QP curve
in Fig.~\ref{munmue})
we find that the right hand side of (\ref{chargedens}) must also be negative. Therefore, at fixed $\mu_n$ 
(say $\mu_n^{[1]}$) the pressure is decreasing if we decrease $\mu_e$ from $\mu_e^{[1]}$ to the lower
value at which the QP is charge neutral. This means
\begin{equation}\label{shown}
p_{QP}(\mu_n^{[1]},\mu_e^{[1]})>p_{QP}(\mu_n^{[1]}).
\end{equation}
Putting (\ref{point1}) and (\ref{shown}) together, we finally obtain (\ref{toshow}). 
This shows that due to the treatment of the phase transition 
as a simplified transition of a one-component system,
the transition densities are shifted to higher values as compared to the ones 
expected from the correct treatment of neutron star matter as a two-component system.

\end{appendix}

\end{document}